%% file: paper.tex
\documentclass[sigplan,screen,nonacm]{acmart}

\usepackage{booktabs}   
\usepackage{subcaption} 

\usepackage{microtype}
\usepackage[nameinlink,capitalise,noabbrev]{cleveref}
\usepackage{listings}
\usepackage[inline,shortlabels]{enumitem}
\usepackage{array}
\usepackage{caption}
\usepackage[normalem]{ulem}
\usepackage{interval}
\usepackage{makecell}
\usepackage{balance}
\usepackage{xspace}
\usepackage{relsize}
\usepackage{microtype}
 
\definecolor{comment}{rgb}{.5,.5,.5}
\definecolor{number}{rgb}{.75,.75,.75}
\DeclareRobustCommand{\textjava}[1]{{\lstset{basicstyle=\footnotesize\ttfamily}\lstinline@#1@}}
\newcolumntype{B}{>{\global\let\currentrowstyle\relax}}
\newcolumntype{^}{>{\currentrowstyle}}

\newcolumntype{C}{>{\bfseries}}

\newcommand\later[1]{}

\newcommand{\cival}[2]{\textcolor{gray}{\relsize{-2}{$\pm$#1}}}
\newcommand{\otoprule}{\midrule[\heavyrulewidth]}

\definecolor{mygreen}{rgb}{0,0.6,0}
\newcommand{\bst}[1]{\textcolor{mygreen}{#1}}
\newcommand{\wst}[1]{\textcolor{orange}{#1}}
\newcommand{\gr}[1]{\textcolor{gray}{#1}}

\newcommand{\impdead}{implicitly dead\xspace}
\newcommand{\lxr}{LXR\xspace}
\newcommand{\zentwo}{Zen 2\xspace}
\newcommand{\zenthree}{Zen 3\xspace}
\newcommand{\coffeelake}{C'Lake\xspace}

\setcopyright{rightsretained}
\acmPrice{}
\acmDOI{10.1145/3519939.3523440}
\acmYear{2022}
\copyrightyear{2022}
\acmSubmissionID{pldi22main-p132-p}
\acmISBN{978-1-4503-9265-5/22/06}
\acmConference[PLDI '22]{Proceedings of the 43rd ACM SIGPLAN International Conference on Programming Language Design and Implementation}{June 13--17, 2022}{San Diego, CA, USA}
\acmBooktitle{Proceedings of the 43rd ACM SIGPLAN International Conference on Programming Language Design and Implementation (PLDI '22), June 13--17, 2022, San Diego, CA, USA}

\begin{document}

\title[Low-Latency, High-Throughput Garbage Collection]{Low-Latency, High-Throughput Garbage Collection (Extended Version)}         


\author{Wenyu Zhao}
\orcid{0000-0002-5148-5736}             
\affiliation{
 \department{School of Computing}             
 \institution{Australian National University}           
  \country{Australia}                    
}
\email{wenyu.zhao@anu.edu.au}          

\author{Stephen M. Blackburn}

\orcid{0000-0001-6632-6001}             
\affiliation{
 \department{School of Computing}             
 \institution{Australian National University}           
  \country{Australia}                    
}
\email{steve.blackburn@anu.edu.au}          

\author{Kathryn S. McKinley}

\orcid{0000-0002-7188-2501}             
\affiliation{
 \institution{Google}           
  \country{United States}                    
}
\email{ksmckinley@google.com}           

\input{abstract}



\maketitle 

\input{intro}
\input{background}
\input{algorithm}
\input{methodology}

\input{results}
\input{discussion}

\begin{acks}                            
This material is based upon work supported by the 
Australian Research Council under project DP190103367.
\end{acks}

\section*{Data Availability Statement}

We made this implementation of \lxr, the DaCapo benchmark suite, and  instructions for reproducing our results~\cite{LXRA:22} and the latest source code~\cite{WZ:22}  publically available.

\balance
\raggedright
\bibliography{venue-long,new}



\end{document}

%% file: abstract.tex
\begin{abstract}
    To achieve  short pauses, state-of-the-art
  concurrent copying collectors such as  C4, Shenandoah, and ZGC use substantially more CPU cycles
  and memory than  simpler collectors. 
  They suffer from
  design limitations:
    \begin {enumerate*}[i)]
    \item  {concurrent copying} 
     with  inherently {expensive read and
      write barriers},
    \item {scalability} limitations due to tracing, and
    \item {immediacy} limitations for mature objects that impose 
      memory overheads.
    \end{enumerate*}

   This paper 
    takes a different approach to
    optimizing responsiveness and throughput. 
    It uses  the
    insight that regular, brief stop-the-world collections  deliver sufficient responsiveness at greater efficiency than concurrent evacuation.
It introduces     \lxr, where stop-the-world collections use reference counting (RC) and
    judicious copying.
     RC  delivers {scalability} and {immediacy}, promptly reclaiming young and mature objects.
     RC, in a hierarchical Immix heap
     structure, 
     reclaims most memory
      {without any copying}.
      Occasional concurrent tracing  identifies cyclic garbage.
      \lxr introduces: {
    \begin {enumerate*}[i)]
    \item {RC remembered sets} for judicious copying of mature objects;
    \item a novel {low-overhead write barrier}
      that combines coalescing reference counting, concurrent tracing,
      and remembered set maintenance;
    \item  {object reclamation while performing a concurrent trace};
    \item {lazy processing of decrements}; and
    \item novel {survival rate triggers} that modulate pause durations.
    \end{enumerate*}}

    { 
    \lxr combines excellent responsiveness and
    throughput, improving over production collectors. 
    On the widely-used Lucene search engine in a tight heap, 
    \lxr delivers 7.8$\times$ better throughput
    and 10$\times$ better 99.99\% tail latency than
    Shenandoah.  On 17 diverse modern workloads in a moderate heap, \lxr outperforms OpenJDK's default G1 on throughput by 4\%  and Shenandoah by 43\%.}
\end{abstract}

%% file: intro.tex
\section{Introduction}
\label{sec:intro}

Modern concurrent garbage collectors are surprisingly expensive. 
Growth in  memory footprints and latency-sensitive applications led vendors to focus on low pause time collectors, such as C4~\cite{TIW:11}, Shenandoah~\cite{FKD+:16}, and ZGC~\cite{Liden:17}.  {While they 
  achieve low pause times, they do so at  memory and CPU costs.  Furthermore, we show that their low pause times do not always translate into low latency for latency-sensitive applications. 
This paper identifies why they are expensive and proposes a very different approach.} We introduce  \lxr (\textbf{L}atency-critical Immi\textbf{X} with \textbf{R}eference counting), 
implement it in OpenJDK, and compare it against these widely-used collectors on diverse contemporary workloads.

The early Garbage First (G1) collector is a copying collector with concurrent tracing~\cite{DFHP:04}. Each collection chooses a set of profitable fixed size regions to evacuate.  It is optimized for throughput and low pause times. C4, Shenandoah, and ZGC built on the G1 foundation, seeking to further reduce pause times, believing lower pause times would translate to improved application latency. These collectors  all  rely on
\begin {enumerate*}[i)]
\item concurrent tracing to identify live objects, 
\item evacuation alone to reclaim space, and
\item expensive read and/or write barriers.
\end{enumerate*}
These choices have fundamental implications on 
\begin {enumerate*}[i)]
\item application (mutator) performance,
\item timeliness of reclamation, and 
\item collector performance.
\end{enumerate*}
They reclaim memory only with concurrent copying, an intrinsically expensive approach that relocates every object 
using  expensive barriers  to maintain coherence of heap references~\cite{PPS:08}. 

\begin{table*}[t]
    \caption{Throughput, latency, and GC pauses on \textsf{lusearch} at a 1.3$\times$ heap.  Short GC pauses \emph{do not} assure low latency.}
    \vspace*{-1.9ex} 
    \input{tables/punchline}
    \label{tab:punchline}
    \vspace*{-1.1ex} 
\end{table*}

{
\cref{tab:punchline} shows the tradeoff Shenandoah makes to achieve low pause times and that those low pause times \emph{do not} translate to low latency on the widely used Lucene enterprise search engine (\textsf{lusearch}). 
It compares G1---the OpenJDK default, optimized for throughput; Shenandoah--optimized for latency; and \lxr---optimized for both, using a tight heap 1.3$\times$ the minimum required by G1 on our \zenthree.  The workload is challenging because it is highly parallel and has a very high allocation rate. We report \emph{throughput} using queries per second (QPS) and total time, and \emph{latency} using query time latency percentiles. }{
Shenandoah has low 50\% and 99\% pause times, but  G1 is lower for the 99.9\% and 99.99\% pauses.  However, low pauses \emph{do not} translate into low application latency, with G1 improving over Shenandoah.  Furthermore, Shenandoah takes a 7$\times$ hit on throughput compared to G1.
Shenandoah cannot keep up with  \textsf{lusearch}'s 9.5\,GB/s allocation rate on this 16-core, 2-way SMT machine.   We also configured Shenandoah to run in a huge heap, 10$\times$ the minimum.  Given this substantial memory headroom, Shenandoah can deliver good throughput and latency.
\lxr delivers  better throughput and latency than G1 and Shenandoah without requiring extra headroom. Although \lxr has slightly longer pauses than G1 and Shenandoah, it improves query latency over both, while also outperforming them on throughput.}

G1, C4, Shenandoah, and ZGC all \textbf{use tracing} 
and must  perform a trace of all live objects to reclaim any memory.  G1 and C4 mitigate this delay with generational collection, reducing the scope of the trace. ({Shenandoah and ZGC have generational variants under development.})  Tracing is unscalable in the limit. Its worst case is a live singly-linked list which defeats parallel tracing and copying~\cite{BP:10}.  \lxr's tracing for cyclic garbage detection also experiences this worse case.  The scalability of reference counting is limited only when a large list dies.
{\lxr achieves immediacy 
and scalability of collection with high performance parallel reference counting 
as its primary mechanism for garbage identification. } 

{C4, Shenandoah and ZGC \textbf{use concurrent evacuation} to optimize pause times, adding high overheads (\cref{tab:punchline}).  They reclaim space only when a region  is  empty: when all objects eventually die or  they evacuate live objects.  Evacuation perturbs the memory hierarchy, using caches and DRAM bandwidth.  Worse, concurrent evacuation requires expensive barriers  to prevent mutator and collector races, which is  intrinsically more expensive than stop-the-world evacuation~\cite{PPS:08}.  {\lxr avoids these overheads with the Immix heap structure~\cite{BM:08} and reference counting with remembered sets, reclaiming most memory without copying, only judicious stop-the-world copying to combat fragmentation.}}

These  collectors \textbf{depend on expensive barriers}, limiting  best-case application performance.   C4, Shenandoah, and ZGC use a load value barrier (LVB)~\cite{TIW:11}, which filters every reference load.   {\lxr does not require a read barrier. It uses a novel low-overhead write barrier for}
\begin {enumerate*}[i)]
\item {concurrent tracing to reclaim cyclic garbage~\cite{Yuasa:90}, for}
\item {building remembered sets, and for}
\item {high performance coalescing reference counting to reclaim both young and old objects promptly~\cite{LP:01,SBYM:13}}.  
\end{enumerate*}

\subsubsection*{Design}

{\lxr's design premise is that regular, brief stop-the-world collections will yield sufficient responsiveness and far greater efficiency than concurrent evacuation.}
\lxr builds on Reference Counting Immix~\cite{SBYM:13} (RCImmix), taking a novel approach to low-latency performance by {limiting concurrency and  copying.} It copies only during stop-the-world pauses. \lxr employs parallelism for scalability in every collection phase. It  exploits existing optimizations, such as  \impdead young objects~\cite{SBYM:13}.    The Immix hierarchical heap divides \emph{blocks} into \emph{lines} and tracks their liveness. 
The application allocates using a fast bump pointer into blocks.  It recycles free lines in a block, when all the objects are on the line are dead.

\lxr uses a field-logging write barrier~\cite{Blackburn:19} with just 1.6\% mutator overhead 
to perform coalescing reference counting,  concurrent tracing, and maintain remembered sets. 
At each pause, \lxr performs reference counting (RC) increments and reclaims free blocks and lines with dead young objects that never receive an increment~\cite{SBYM:13}. It judiciously  copies young  surviving objects to defragment blocks~\cite{SBYM:13}. \lxr triggers occasional concurrent tracing using Yuasa's Snapshot At The Beginning (SATB) tracing algorithm~\cite{Yuasa:90} to collect old dead cycles and  objects with stuck reference counts.  SATB tracing is elegantly implemented 
with the same write barrier used for coalescing reference counting~\cite{AP:03}. 
 We show how to soundly delete objects while an SATB trace is in progress, such that the trace may span multiple RC epochs. At the beginning of an SATB trace, \lxr identifies candidate \emph{evacuation sets} of blocks with high fragmentation. The trace initializes each remembered set and the write barrier keeps them up to date. After the trace completes, \lxr reclaims free blocks and lines at the next RC pause and evacuates any still fragmented blocks in the evacuation sets.

\lxr modulates  pause times using three novel approaches.  \begin{enumerate*}[i)] \item Collection triggers use survival rate prediction to control the expected work \lxr performs in each RC epoch to reclaim dead young objects and to determine when SATB traces are likely to be profitable.  \item Lazy concurrent decrements reclaim old dead objects. \item Incremental copying of one or more evacuation sets defragments blocks with old objects. \end{enumerate*}

 Techniques novel to LXR are: \begin{enumerate*}[i)] 
    \item SATB tracing spanning multiple RC pauses, 
    \item combining RC and remembered sets,  
    \item a single, simple write barrier that combines coalescing reference counting, concurrent tracing, and  remembered sets for copying with RC, \item survival rate collection triggers,  and
    \item lazy concurrent decrements.
\end{enumerate*}

\subsubsection*{Implementation}

{
We implement \lxr in MMTk~\cite{BCM:04a,BCM:04b,MMTk:22} on OpenJDK 11 and compare against G1, Shenandoah, and ZGC on a range of modern workloads.   We demonstrate that this fresh approach to  latency-sensitive collection achieves latency results that improve over these production collectors without additional hardware requirements.  \lxr delivers lower latency than Shenandoah and ZGC on each of the four latency-critical benchmarks we evaluate, and does so while outperforming G1 on throughput, while Shenandoah suffers 77\% and 37\% slowdowns on 1.3$\times$ and 2$\times$ heaps respectively.  We are optimistic that this new design approach will lead to more efficient latency-sensitive collectors and a new class of low-overhead throughput-oriented collectors.
}


%% file: tables/punchline.tex
{
\setlength{\tabcolsep}{.7ex}
\sffamily

\begin{tabular}{BCrrrrr^r^r^r^r^r^r^r^r^r}
    \toprule
                                 &  & \multicolumn{2}{c}{\textbf{Throughput}} &           & \multicolumn{4}{c}{\textbf{Query Latency} (ms)} & \hspace*{2em} & \multicolumn{4}{c}{\textbf{GC Pauses} (ms)}                                                                             \\
    Algorithm                    &  & QPS                                     & Time (s)  &                                                 & 50\%          & 99\%                                        & 99.9\%    & 99.99\%    &  & 50\%      & 99\%      & 99.9\%    & 99.99\%   \\
    \cmidrule{1-1}\cmidrule{3-4}\cmidrule{6-9} \cmidrule{11-14}
    G1                           &  & 112\,K                                  & 4.7       &                                                 & \bst{0.1}     & 12.0                                        & 14.6      & 15.9       &  & 0.4       & 0.9       & \bst{1.1} & \bst{1.2} \\
    Shenandoah                   &  & 15\,K                                   & 34.5      &                                                 & 0.3           & 78.0                                        & 116.1     & 127.8      &  & \bst{0.1} & \bst{0.3} & 2.2       & 3.0       \\

    \lxr                         &  & \bst{119\,K}                            & \bst{4.4} &                                                 & \bst{0.1}     & \bst{3.0}                                   & \bst{8.0} & \bst{13.1} &  & 0.9       & 1.4       & 1.8       & 2.9       \\
    \cmidrule{1-1}\cmidrule{3-4}\cmidrule{6-9} \cmidrule{11-14}
    \gr{Shenandoah$^{10\times}$} &  & \gr{170\,K}                             & \gr{3.1}  &                                                 & \gr{0.1}      & \gr{10.8}                                   & \gr{14.4} & \gr{16.6}  &  & \gr{0.2}  & \gr{0.6}  & \gr{0.8}  & \gr{0.9}  \\
    \bottomrule
\end{tabular}
}

%% file: background.tex
\section{Background and Related Work}
\label{sec:background}

This paper exposes problems with state-of-the-art concurrent collector design and develops \lxr, a novel collector design to solve those problems.  We begin with the background that underpins \lxr and prior collectors.

\subsection{Reference Counting}
\label{sec:rc}

Reference counting (RC) tracks the number of references to each object, reclaiming objects with no references.  Traditional RC performs increments for every new reference to an object and decrements for every destroyed reference~\cite{Collins:60}.  Because pointers change frequently, 
this approach is too inefficient for performance-critical applications.   Temporal coarsening 
almost entirely removes this overhead.  The underlying insight is that for any period $t_n \to t_{n+1}$, it is sufficient to identify all modified pointer fields, $p$,  and apply a decrement to the referent of $p$ at $t_n$, and an increment to the referent of $p$ at $t_{n+1}$.  
This insight underpins both \emph{deferral} for highly mutated stack variables~\cite{DB:76}, and \emph{coalescing} for heap variables~\cite{LP:01}. \lxr applies
\begin{description}
    \item[\emph{deferral}] by capturing all root pointers during brief mutator stop-the-world pauses (epochs) at $t_n$ and $t_{n+1}$~\cite{BAL+:01} and
    \item[\emph{coalescing}] by capturing the to-be-overwritten referent the first time a heap pointer is written in the period $t_n \to t_{n+1}$, and the location of the pointer to establish its referent at $t_{n+1}$, ignoring  intermediate referents~\cite{LP:01}.
\end{description}
Every overwritten referent in the before image ($t_n$) receives a decrement.  Every fresh referent in the after image ($t_{n+1}$) receives an increment.
%
%
%
For root pointers, \lxr implements deferral by applying an increment to all root-reachable objects at $t_n$ and remembering them in a buffer so that it can apply a matching decrement at $t_{n+1}$~\cite{BAL+:01}. 
Prior to coalescing~\cite{LP:01}, applying RC to heap objects required expensive synchronization and was considered an impediment to adoption in parallel systems. Coalescing significantly reduces the synchronization work of RC systems.  Coalescing with buffering also opened up optimizations such as parallel processing of the increment and decrements.
Further optimizations and using a hierarchical mark-region (Immix) heap have delivered reference counting collectors that can outperform the fastest tracing collectors~\cite{AP:03,BM:03,SBF:12,SBYM:13}.  

\begin{figure}[t] 
\centering
        \includegraphics[width=\columnwidth]{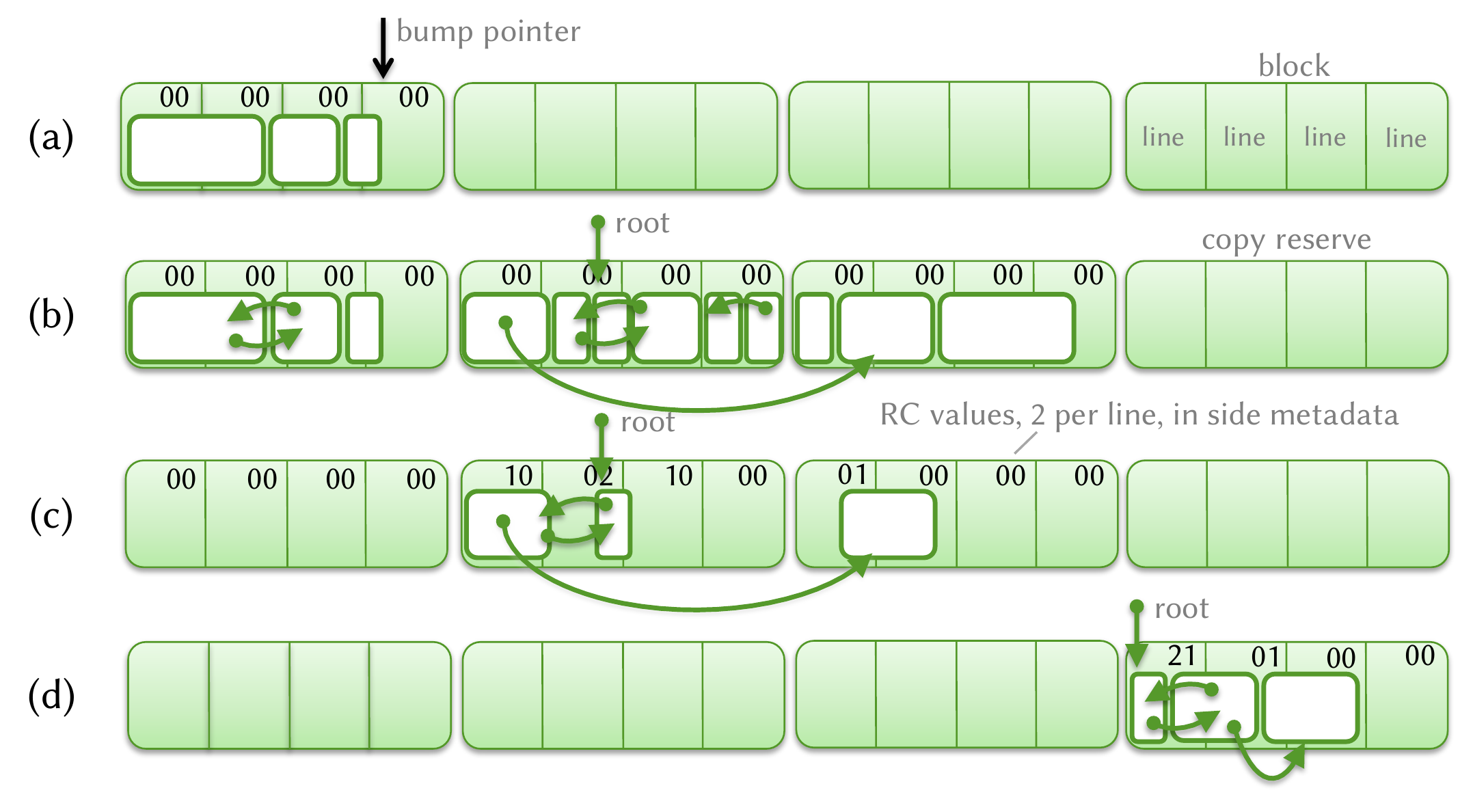}
\vspace*{-4ex}
\caption{RC in an Immix heap with 2 bit RC values stored in line metadata: 0 is free, 1 is 1  RC to the object starting here, 2 is 2 RCs, and 3 is 3 or more RCs, simplified to a maximum of two objects per line, ignoring alignment. (a) Immix heap allocation. (b) Object graph before collection. (c) After collection.    (d) The \impdead optimization evacuates young objects, creating 3 free blocks instead of 2.} \label{fig:lxr-heap}
\end{figure}

\lxr uses the extremely effective \emph{\impdead} optimization which applies temporal RC coarsening to young objects~\cite{SBF:12}. First, young objects allocated since $t_n$ never generate decrements in the period $t_n$ to $t_{n+1}$ since the initial state of their pointer fields is null.  Second, they cannot receive any decrements from the heap or stack since, by definition, there were no pointers to them at $t_n$.   Only their pointer state at $t_{n+1}$ is thus relevant and it will only create increments.   Young objects that receive no increments at $t_{n+1}$ are dead and most do die, following the weak generational hypothesis~\cite{Ungar:84,LH:83}. It is therefore correct to 
\begin{enumerate*}[i)]
    \item ignore impossible decrements from young objects, and
    \item generate increments from young objects only if and when the object receives its first increment at $t_{n+1}$.
\end{enumerate*}
A coalescing write barrier thus correctly ignores mutations to young objects.

Since the first GC after allocation will establish \emph{all} references to surviving young objects, they can be correctly copied~\cite{SBYM:13}. \cref{fig:lxr-heap}  illustrates these optimizations (see~\cref{sec:immix}). \cref{fig:lxr-heap}(a) and (b) show that the initial RC values are zero 
in the young object metadata. \cref{fig:lxr-heap}(c) shows the RC values after increments for each root reachable young object. Note that the first block of only dead young objects is free { (some in cycles)}, no RC was ever changed, and \lxr can reclaim the block before processing decrements. \cref{fig:lxr-heap}(d) shows combining copying live objects while establishing their RC values, efficiently produces more free blocks.

Reference counting is fundamentally a local property. The liveness of an object can usually be determined without a global operation such as a heap trace and increases the immediacy of reclamation.
RC increment and decrement processing grows only as a function of the rate of graph mutations and are both easily parallelized in general. However, if the head of a singly-linked list dies, recursively decrementing the list defeats 
parallelism. \lxr limits the impact of this case by performing decrements concurrently. In contrast, tracing parallelism is defeated by a singly-linked list every time it is traced. We observe this worst case degrading G1 and Shenandoah throughput compared to \lxr in \cref{sec:res-throughput}.

{ Reference counting is incomplete because it cannot identify cycles of garbage and previously was considered incompatible with copying, lacking global information for identifying and redirecting references to relocated objects.  \lxr reclaims dead cycles via a backup trace.  
 \lxr uses the \impdead optimization to copy young objects. 
\lxr adds novel techniques to combine the immediacy of RC with the defragmentation and locality benefits of  evacuation for mature objects.} We introduce remembered sets for RC and use them to combine block evacuation with RC for the first time. 

\subsection{Read and Write Barriers}
\label{sec:barriers}

Read and write barrier code 
intercepts mutator reads and writes to enforce invariants and monitor object graph mutations~\cite{YBFH:12}.  Coalescing reference counting uses a write barrier that, upon the first update to a pointer field, enqueues the overwritten referent for a future decrement and remembers the address of the updated field so that at the next collection, the eventual referent held at that address can receive an increment~\cite{LP:01}. \citet{AP:03} observed that this barrier may also implement SATB tracing~\cite{Yuasa:90}, which we exploit in \lxr.  The barrier is easily implemented at the object or the field granularity~\cite{Blackburn:19}.  \cref{sec:res-overheads}  shows the field-logging barrier used by \lxr adds 1.6\% overhead.  G1~\cite{DFHP:04} uses several write barriers that it enables and disables during different phases of the collection.  C4~\cite{TIW:11}, Shenandoah~\cite{FKD+:16}, and ZGC~\cite{Liden:17} all use the loaded value barrier (LVB), which filters every reference load, clearing bits if necessary, and conditionally takes further action.   Shenandoah has an additional SATB write barrier for concurrent marking.  Because applications typically execute field loads an order of magnitude more frequently than field stores (e.g.\ 64.3/$\mu$s v 4.3/$\mu$s), read barriers, such as the LVB, are  on the order of five times more expensive than an object remembering barrier~\cite{YBFH:12}. 

\subsection{Concurrent Tracing}
\label{sec:satb}

Concurrent tracing algorithms identify all live objects concurrently with mutators, which may also be modifying the object graph~\cite{DLM+:78,Steele:75,Yuasa:90}.  
{G1, Shenandoah, and \lxr use \citeauthor{Yuasa:90}'s Snapshot At The Beginning (SATB) algorithm~\cite{DFHP:04,FKD+:16, Yuasa:90}. 
SATB exploits the observation that once an object is unreachable, it  remains unreachable, and  it is therefore correct to perform a trace over a stale heap snapshot.   SATB captures program roots at the start of the trace. It intercepts the first time any pointer is overwritten during the trace and  traces  the about to be destroyed reference, along with all objects created since the start of the trace.}  \lxr combines the SATB trace with the \impdead optimization, such that the trace only considers objects that survive an RC collection, significantly reducing the scope of its trace. 

\subsection{Concurrent Copying}
\label{sec:conccopy}

As well as ensuring liveness is correctly established despite a concurrently mutated object graph,  concurrent copying collectors (C4, Shenandoah, and ZGC) 
incur additional complexity in order to maintain referential integrity dynamically, when mutators dereference pointers which refer to objects concurrently being moved.  The literature contains many such algorithms, including \citeauthor{Brooks:84}'s algorithm~\cite{Brooks:84},  initially used by Shenandoah~\cite{FKD+:16}, and \citeauthor{Baker:78}'s algorithm~\cite{Baker:78}, which is the basis for the loaded value barrier (LVB) used
used  by C4~\cite{TIW:11}, ZGC~\cite{Liden:17}, and recent versions of Shenandoah~\cite{FKD+:16}.  Another approach is page protection and double mapping to perform concurrent evacuation while capturing any stale references~\cite{AL:91,CTW:05,KP:06,TIW:11}.  \citeauthor{PPS:08} evaluated a range of concurrent copying collectors and found that the best of them incurred a throughput and latency overhead of 20\%~\cite{PPS:08}. 

\lxr's design is based on the observation that \emph{concurrent copying is intrinsically expensive}. 
   \lxr instead 
   minimizes copying and
   eliminates concurrent copying completely by copying only during mutator pauses.

\subsection{Region-Based Evacuating Collectors}
\label{sec:g1}

G1, C4, Shenandoah,  and ZGC are region-based and strictly-copying~\cite{ZB:20}.    They partition the heap into fixed size regions for independent collection (e.g.,\ G1 uses 1--32\,MB depending on heap size). 
Prior to each GC cycle, they use heuristics such as age and fragmentation to select regions to collect, evacuating any live objects as they trace to another region.   Concurrent tracing identifies garbage and fixes pointers to moved objects. Because  tracing is their primary mechanism for reclamation, they are subject to the inherent scalability limitations suffered by all strictly-tracing collectors~\cite{BP:10}. As the heap grows, the collector is not guaranteed  to trace the entire heap in a timely manner, e.g., because of lack of parallelism in the heap graph or when parallel tracing threads compete with the mutator. 


G1 uses a write barrier and remembered sets to identify pointers to evacuated objects, redirecting them as it moves objects during a mutator pause, which can cause long pauses~\cite{DFHP:04}. C4, ZGC, and Shenandoah all target low ($\leq$10\,msec) pauses, necessitating concurrent evacuation.  They use a loaded value barrier (LVB) on every pointer load to maintain referential integrity and evacuate concurrently with the mutator~\cite{TIW:11}.  As \cref{tab:punchline} highlights, their designs impose substantial costs on the mutator. C4 uses page protection to enforce referential integrity while lazily forwarding pointers to moved objects~\cite{CTW:05,TIW:11}.
{These algorithms suffer because they
\begin{enumerate*}[i)] 
    \item \emph{exclusively trace} to identify garbage, limiting scalability and timeliness, 
    \item \emph{exclusively copy} to reclaim space, 
    \item \emph{use expensive read and write barriers}, 
    \item \emph{evacuate concurrently} to meet low pause time targets, and
    \item \emph{have long collection cycle times} from concurrent whole-heap tracing, leading to relatively large memory requirements.   
\end{enumerate*}}

\subsection{Immix Hierarchical Heap}
\label{sec:immix}

\lxr builds on RC-Immix~\cite{SBYM:13} and its heap structure. 
Immix and RC-Immix combine the locality and simplicity of evacuating collectors without a strict copying requirement~\cite{BM:08,SBYM:13}.  Immix structures the heap in a hierarchy of 32\,KB blocks, which are composed of 256\,B lines. As shown in~\cref{fig:lxr-heap}(a),  thread-local allocators use a bump pointer to allocate into a block. Objects may span lines but not blocks.  Allocators reuse partially free blocks by skipping over live lines and allocating into free lines. For example the last two lines in the third block in \cref{fig:lxr-heap}(c) are candidates for allocation after a collection. Allocators acquire a new free or partially-free block from a global pool when they exhaust their current block. Collectors may trace~\cite{BM:08} or reference count~\cite{SBYM:13} the Immix heap, identifying completely free lines and blocks. 

The RC-Immix collector tracks the number of live objects on a line and a block. \lxr uses a different approach, storing RC bits for each object (see \cref{sec:alg-identrc}). \cref{fig:lxr-heap}(c) shows these RCs, abstracted for illustrative purposes to two per line, at the end of a reference counting epoch, when no object is evacuated. \lxr reclaims old objects using reference counting without moving the live objects.  Immix collectors may opportunistically copy objects to reclaim highly fragmented blocks.  \cref{fig:lxr-heap}(d) shows how \lxr uses the \impdead optimization to evacuate young objects, copying them into a free block, partially free blocks are preferred. Evacuation eliminates the fragmentation shown in~\cref{fig:lxr-heap}(c).  
\lxr also targets blocks with old objects and very few live lines to evacuate during cycle tracing, producing completely free blocks at low cost, similar to Immix (see \cref{sec:alg-reclamation}).

\subsection{Orthogonal Optimizations}
\label{sec:opts}

{Many optimizations target reducing collection and pause times, such as escape analysis~\cite{PG:92,CGS+:03}, concurrent stack scanning~\cite{YNKY:02,SKK+:05,KPS:09,Osterlund:20,Kennke:21}, barrier elision~\cite{VB:04}, 
compressed pointers~\cite{LA:05}, and large object spaces. 
They are orthogonal to our approach and we do not evaluate them here.}






%% file: algorithm.tex
\section{The \lxr Algorithm}
\label{sec:algo}

\begin{figure*}
    \includegraphics[width=\textwidth]{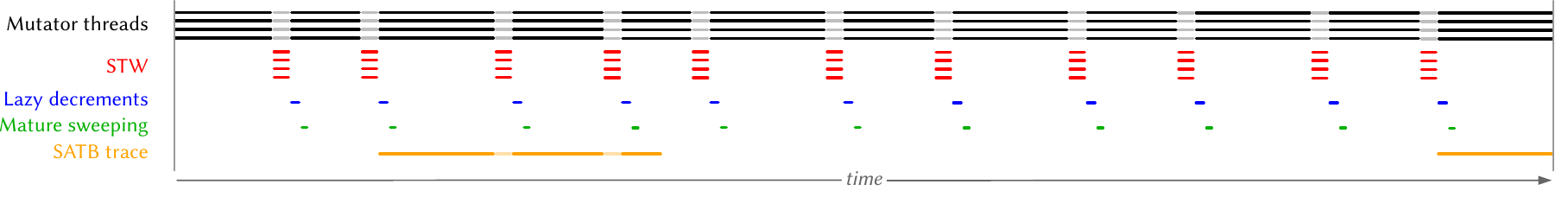}
\vspace*{-2em}
    \caption{A simplified LXR timeline with four mutator threads. Four parallel STW GC threads increment object RCs, perform evacuation, and sweep young blocks. A concurrent LXR thread performs lazy decrements, old block sweeping, and the SATB trace. Unshown mutator activities: allocation and write barriers. 
We show 11 RC epochs,  a complete and partial SATB trace.  
}

    \label{fig:timeline}
\end{figure*}

A key design insight for \lxr is that \emph{{regular brief STW pauses provide a highly-efficient, low-latency approach to garbage collection}}. We next present \lxr's algorithms and optimizations for {allocation}, {identification}, and {reclamation}. \cref{fig:timeline} overviews \lxr's timeline of stop-the-world (STW) and concurrent collection activities.

\subsection{Allocation and Heap Structure}
\label{sec:alg-alloc}

\lxr uses the Immix hierarchical heap structure and a bump point allocator. 
It uses two global lock-free data structures to maintain lists of free and partially free blocks respectively. Thread local allocators obtain blocks from these lists.  Following Immix, \lxr uses the partially free blocks first to maximize the availability of free blocks for large allocations which are managed separately.

When allocating into a partially free block, \lxr uses a \emph{reference count table} (\cref{sec:alg-indentification}), to identify unavailable lines, skipping over them to find free lines.   Because objects may straddle lines but not blocks, following Immix~\cite{BM:08}, 
\lxr conservatively assumes that the first line after a used line is unavailable,  side-stepping computations to track objects that straddle two lines.  When \lxr applies the first RC increment to an object, if the object is larger than a line, it writes a non-zero value into the reference count table entry for each trailing line except the last, to ensure  trailing lines are not reused.  The last line does not need an entry  because the conservative treatment of straddling objects already accounts for it correctly, as shown in block 3 of \cref{fig:lxr-heap}(c).

\lxr employs the dynamic overflow optimization from Immix to fill lines in partially free blocks.  If there are one or more free lines between the cursor and the limit, and the current object does not fit, i.e., it is bigger than a line, then we allocate that object in an initially completely free overflow block for medium objects. Dynamic overflow avoids wasting lines.
In the common case, allocation is  fast and the allocator provides good spatial locality for contemporaneously allocated objects. Objects larger than half a block in size (16\,KB) are delegated to a large object allocator.


If the runtime requires that \lxr zero memory before use, it zeros each free region immediately before allocating into it. Thus, \lxr zeroes free blocks in bulk~\cite{YBFSM:11}. For a partially free block, it zeros contiguous free lines immediately prior to allocating into them.  Most managed languages require memory to be zeroed.  OpenJDK optionally provides compiler-injected per-allocation zeroing.  In this case, \lxr does not need to perform zeroing at allocation time.

\subsection{Identification}
\label{sec:alg-indentification}

\lxr promptly identifies dead objects using reference counting with deferral~\cite{DB:76}, coalescing~\cite{LP:01}, and \impdead~\cite{SBF:12} optimizations, with a back-up concurrent trace~\cite{Yuasa:90} to collect dead cycles and dead objects with stuck reference counts.

\lxr uses regular, brief stop-the-world (STW) pauses to coherently scan runtime roots including mutator threads and stacks.
Compared to taking a more aggressive on-the-fly approach to almost entirely remove pauses~\cite{LP:01}, 
\lxr instead uses STW pauses as opportunity for highly efficient, targeted evacuation (\cref{sec:alg-reclamation}) and increment processing.

%
%

\subsubsection{RC Identification}
\label{sec:alg-identrc}

Each RC epoch applies all increments first and then decrements. For each store to a reference field during mutator execution, \lxr's write barrier (\cref{sec:alg-barrier}) puts the overwritten reference in a \emph{decrement buffer} and puts the address of the updated field in a \emph{modified fields buffer}. Each RC epoch, \lxr performs root scanning and performs an increment for each root-reachable object and buffers a corresponding decrement for the next epoch.  \lxr processes the modified fields buffer by applying an increment to the current referent of each modified field.  After applying all increments, it processes the decrement buffer, which includes the root targets from the prior epoch and all overwritten references from the last mutator epoch, applying a decrement to each target object. 

Whenever a decrement transitions a count from $1 \to 0$, \lxr enqueues the dead object for recursive decrements, since all pointers in the dead object are about to be destroyed. 
{\lxr processes the recursive decrement queue by scanning each object and applying a decrement to each of its referents.} Once the  decrement queue and the recursive decrement queue are exhausted, the RC epoch is complete.
In principle, recursive decrement processing could apply decrements to most objects,  leading to arbitrarily long pauses, e.g., when the last reference to a large object graph is destroyed.  
We address this limitation, which is common to all RC algorithms, by restarting the mutators and performing concurrent lazy decrement processing (\cref{sec:alg-lazydec}).

\lxr is implemented in OpenJDK, which lacks sufficient object header bits for a reference count.   \lxr therefore stores reference counts in a \emph{reference count table}, where each count is reachable from an object address via simple address arithmetic. 
Given a minimum object size of 16 bytes, we allocate an $N$-bit reference count for each 16 bytes of allocated memory.  \citet{SBF:12} found 
$N_{rc} = 3$ was sufficient.   
We select two bits for simple address arithmetic and to limit memory overhead.  
With $N_{rc} = 2$, each 256\,B line of allocated data consumes four bytes of metadata.  The \lxr allocator (\cref{sec:alg-alloc}) scans these densely packed reference counts. Free lines and blocks have all zero counts.

{
 If a reference count for an object reaches 3, then we \emph{stick} its count, and do not apply further decrements or increments to it. We rely on the concurrent trace to identify dead objects with stuck reference counts.   In our analysis, 16 of 17 benchmarks  have very few stuck counts  (0 to 1\% in \cref{tab:analysis}).} \textsf{Batik} is an outlier with 6.6\%.




\subsubsection*{Optimization: Lazy Decrements}
\label{sec:alg-lazydec}

\lxr processes RC decrements lazily, concurrent with the mutator.   
Because \lxr reclaims a lot of memory, i.e., all the memory consumed by dead young objects~\cite{SBF:12}, after applying RC increments and \emph{without} processing any decrements (\cref{sec:rc}), lazy processing of decrements affects the promptness only of mature object reclamation. 
 Lazy decrements are processed immediately after each brief STW pause with higher priority than SATB trace operations, thus the impact on prompt mature object reclamation is small.  Lazy decrement processing has the further advantage of addressing the problem of recursive decrements taking arbitrarily long.  
If the next RC epoch starts and \lxr still has decrements to process, it finishes them first. {In our analysis, among 17 benchmarks only \textsf{xalan} (21\%), \textsf{lusearch} (5\%), and \textsf{avrora} (1\%) have pauses that must process unfinished lazy decrements (\cref{tab:analysis}).}

\subsubsection*{Heuristic: RC Triggers}
\label{sec:alg-rctriggers}

%
%
\lxr triggers an RC pause if the heap is full or if increment or survival thresholds are reached.   The \emph{increment threshold} protects against long pauses due to increment processing.   The \emph{survival threshold} uses a prediction of young object survival rates to protect against long pauses due to recursive increments and copying of young objects.  {By using a predictor, the trigger targets controlling the \emph{expected} case RC pause time, favoring throughput, in contrast to controlling worst case RC pause times with  an allocation trigger.}  We use an exponential decay
 predictor,  conservatively biasing it to higher survival rates.  After each RC, if the survival rate is higher than predicted, {then the new prediction uses  3/4 of the latest observation and 1/4 of the old prediction.  Otherwise we reverse the weights,  1/4 (latest) :  3/4 (previous).}   Although  survival rates vary greatly between workloads (see \cref{tab:stats}), this predictor is  very effective.

\subsubsection{SATB Identification}
\label{sec:alg-identsatb}

\lxr periodically uses \citeauthor{Yuasa:90}'s SATB concurrent tracing algorithm~\cite{Yuasa:90} to collect dead objects RC cannot identify: objects in cycles and objects with stuck reference counts.   SATB is an elegant match to coalescing RC since both depend on capturing the referents of overwritten pointer fields starting at some time $t$. RC enqueues the overwritten referents so that it can later decrement their reference counts, while SATB must include the overwritten referents in its trace, since they comprise part of the initial heap snapshot at time $t$.   

An SATB tracing epoch  starts during a regular RC collection pause, seeded with the root set from the RC collection. The trace continues concurrently with the mutator, adding additional edges as they are provided by the write barrier.  \cref{sec:alg-barrier} describes how SATB piggybacks on the RC write barrier with no additional mutator overhead.

\subsubsection*{Optimization: SATB with Interruptions}
\label{sec:alg-rcsatb}

~Note that it is correct to add new referents to the SATB trace.   We exploit this property to allow, {as needed}, a single SATB trace to span multiple RC epochs, $t_n$ to $t_{n+k}$, giving \lxr the freedom to perform RC collections as frequently as necessary, independent of the time taken to perform a full heap trace. 
SATB tracing time is governed by the size and shape of the reachable object graph and may not be amenable to parallelism~\cite{BP:10}.  Since the SATB trace must visit all objects that were reachable at the \emph{start} of the trace, 
{RC  deleting objects } during the SATB trace presents a challenge to correctness.  However, deferring RC until {the trace completes would be} a barrier to timely reclamation.   

We generalize SATB tracing to handle \emph{interruptions} (RC collections) during the trace.   The correctness of SATB depends on visiting all objects reachable at the time the SATB began. 
We note that we can achieve this goal by enforcing the invariant that  RC may never delete an unmarked object while an SATB trace is underway.  We implement this invariant {with mark bits stored in side metadata} and by immediately marking and scanning any mature object that RC determines is dead if the SATB has not already marked it.  Note that RC will reclaim space promptly regardless of whether the SATB mark bit is set.  However, once the object's mark bit 
is set, the SATB will not attempt to visit the object, avoiding the possibility of the SATB following a reference to a deleted object.  These potential references come  both from objects no longer reachable and references held in the SATB's mark stack. 
The collector clears these mark bits only after the SATB epoch is finished.  Reclamation of unmarked objects happens during the first RC epoch that starts after the SATB finishes.

\subsubsection*{Optimization: Mature-Only SATB}
\label{sec:alg-matsatb}

{\citeauthor{Yuasa:90}'s SATB algorithm conservatively keeps  new objects live. 
  Using the \impdead optimization, we  correctly apply SATB tracing only to mature objects that survive an RC collection.} 
\lxr implements this optimization by ignoring objects with a zero reference count when it performs the SATB trace. { This optimization substantially reduces the SATB's working set and  eliminates its  prior conservative treatment of objects allocated during the SATB trace.} 

\subsubsection*{Heuristic: SATB triggers}
\label{sec:alg-satbtrig}
%
%

SATB traces place a burden on the mutator, but are essential to defragmenting mature objects and freeing cyclic garbage and objects with stuck counts.    To maximize their effectiveness and limit their costs, we trigger SATB traces judiciously.  \lxr has an available free block trigger and a predicted heap \emph{wastage} trigger for SATB. We define wastage as uncollected dead mature objects and fragmentation.  If an RC yields fewer clean blocks than a prescribed threshold, the next pause will initiate an SATB trace.  We also trigger an SATB if predicted wastage reaches a threshold percentage of the heap. Each RC epoch, we use a live block predictor to predict the number of live blocks that will be in the heap if we perform an SATB trace.  This predictor is driven by live block observations after each SATB trace.  Similar to the RC trigger, the predictor uses a 1:3 ratio conservatively  biased exponential decay.   We use the predictor to estimate wastage after each RC by taking the difference between the current live blocks and the predicted live blocks.  




\subsection{Reclamation}
\label{sec:alg-reclamation}

At its simplest, \lxr is a non-moving reference counting collector.  We augment this algorithm with reclamation performed as a result of the SATB trace and two optimizations that opportunistically copy objects to defragment the heap and improve allocation locality.

\subsubsection{RC Reclamation}
\label{sec:alg-recrc}

\lxr reclaims objects with a reference count of zero after applying all increments.   Because young objects 
start with zero reference counts and die quickly, fully and partially free blocks with young objects are common.  \lxr reclaims free lines and blocks by sweeping the blocks that contain young objects and examining their reference count table.  If all counts are zero, then the entire block is free and \lxr returns the block to the global free block list, where it is available for use by any part of the collector, including large object allocation.   Otherwise, if lines are free in a block, it places the partially free block on the global partially free block list.


\subsubsection*{Optimization: Lazy Reclamation}
\label{sec:alg-reclazymature}


After sweeping blocks containing young objects, \lxr resumes the mutator threads and   processes decrements concurrently.  As soon as it finishes all the decrements, \lxr selectively sweeps those blocks containing objects which received a decrement, returning any partially or completely free blocks to the global allocator.

\subsubsection{SATB Reclamation}
\label{sec:alg-recsatb}

At the start of an RC epoch, \lxr checks whether an SATB trace has completed. 
If a trace is complete, SATB-identified dead objects will be unmarked, but will have a non zero reference count. Note these objects cannot have outstanding RC increments since the SATB algorithm guarantees that they were dead at the start of the SATB, which was at least one epoch earlier.
When \lxr finishes its increments and decrements, the decrement sweep will include  clearing the reference counts of all dead unmarked objects from the SATB trace and will reclaim objects dead due to decrements and dead due to SATB tracing. 

\subsubsection*{Optimization: Young Evacuation}
\label{sec:alg-recyoung}

\citet{SBYM:13} observe that since all surviving young objects will receive an increment from every referent during their first collection, the first collection may move them, safely redirecting (forwarding) each of the referents in the process.  Furthermore, they show that the Immix heap structure allows this movement to be performed \emph{opportunistically} on a per-object basis, according to space and time remaining.  We apply this optimization to \lxr and note that it can be slightly improved by observing that young objects can be trivially identified as any object that receives a $0 \to 1$ increment.

\subsubsection*{Heuristic: All Young Evacuation} In our implementation, we mark the blocks which only contain young objects, i.e., these blocks are empty when assigned to a thread-local allocator prior to this RC epoch. We then target all these blocks for young object evacuation. This optimization exploits the generational hypothesis. Since few objects survive, \lxr will copy only a few objects, creating completely free blocks without much copying, and creating locality by compacting the survivors into the partially free blocks with other survivors from this collection and prior collections.  It will copy these survivors into completely free blocks when the partially free blocks are exhausted or when it applies dynamic overflow for medium size objects~\cite{BM:08}. \cref{fig:lxr-heap}(d) shows this optimization when all the blocks contain only young objects. {If there are no free or partially free blocks, it can stop copying young objects and increment their reference counts in place.}

\subsubsection*{Optimization: Mature Evacuation}
\label{sec:alg-recmature}

\citeauthor{SBYM:13} note that  copying and RC can be combined during a stop-the-world full heap trace for collecting cycles~\cite{SBYM:13}.   This opportunity is not available to \lxr since it does not perform a stop-the-world full heap trace.  
We observe that the write barrier \lxr uses for RC and SATB (\cref{sec:alg-barrier}) is sufficient  to maintain remembered sets, allowing portions of the heap to be independently evacuated during RC collection pauses.  Each \lxr remembered set tracks all pointers into an \emph{evacuation set}.  An evacuation set is a set of blocks within a contiguous region of the heap that will be evacuated together.  \lxr uses either 4\,MB regions, leading to many remembered sets, or the whole heap, yielding a single remembered set.   Prior to each SATB, \lxr identifies and marks \emph{target blocks}, which will form the evacuation set(s).  We use the reference count table as an upper bound on the number of live bytes in each block. For all blocks that have less than 50\% occupancy, we sort them from the lowest to highest occupancy, and mark the $N$ lowest occupancy blocks as \emph{target blocks}.

The remembered sets are not maintained continuously, but only from the start of the {SATB to the RC pause in which the set is evacuated}.  Each set is bootstrapped by piggybacking on the first SATB trace, which must traverse every pointer into the evacuation set.   Each such pointer is appended to the region's remembered set.  The remembered set is kept up to date until the  set is evacuated, via the write barrier, 
which remembers each subsequently-created reference into the evacuation set.  

If remembered sets need to be maintained for more than one RC epoch, entries can become stale, leading to a correctness concern.   Each remembered set entry is a pointer to the location of an incoming reference.  It is benign for the reference to be overwritten, however if the source object dies and the space is reused, the remset entry may point to a non-pointer.   In that case, it could be possible for a non-pointer value to be treated as a pointer, and in the unlikely event that the value appeared to point to an object being evacuated, the collector would incorrectly update it. 
We avoid this problem by maintaining a reuse counter for each line which is reset at each SATB.  Each remset entry is tagged with the reuse count for the source line.  If at evacuation time the source line is newer than the remset entry, we discard the entry.

\lxr evacuates each set by tracing through the blocks using the current collection roots and the remembered set as roots.   When the evacuation encounters an object in the evacuation set, it copies the object into a destination in a free or partially free block, updates the incoming reference, and leaves a forwarding pointer.   If the object has already been copied, the incoming pointer is updated using the forwarding pointer stored in the stale object.  When the trace encounters a pointer outside the evacuation set, it ignores it, limiting the scope of the evacuation to within the evacuation set.
At each STW pause, \lxr evacuates one or more evacuation sets and may use a time budget. 

\begin{figure}
\begin{lstlisting}
void putfield(ref* field, ref value) {
  if (isUnlogged(field))      /* non-atomic    */
    logField(field);
  *field = value;             /* store value   */
}
void logField(ref* field) {
  if (attemptToLog(field)) {  /* synchronized  */
    ref old = *field;         /* remember old  */
    markAsLogged(field);
    decbuf.push(old);         /* log old value */
    modbuf.push(field);       /* log address   */
  }
}
\end{lstlisting}
 \vspace*{-3.2ex}
\caption{\lxr's field write barrier has 1.6\% overhead.} 
    \vspace*{-2.1ex}
\label{fig:barrier}
\end{figure}

\subsection{Write Barrier}
\label{sec:alg-barrier}

For each field the mutator overwrites during $t_n \to t_{n+1}$, \lxr needs to know the field's value at $t_n$ and at $t_{n+1}$.   It uses the former for SATB tracing and to generate decrements for destroyed references. It uses the latter to create remembered set entries and  increments for new references. The barrier may operate at one of two granularities. It can remember \emph{objects} containing fields that are overwritten or with slightly higher mutator overhead, but greater precision, it can remember just overwritten \emph{fields}~\cite{Blackburn:19,LP:01,SBF:12}. We implemented both, but our results use the field barrier shown in \cref{fig:barrier}.


LXR's field-remembering barrier uses one unlogged bit per field. 
LXR uses side metadata to hold the unlogged bit used by both \textjava{isUnlogged()} and \textjava{logField()}.   At each pointer store, the barrier checks the field's unlogged bit. If the field is unlogged, the slow path adds the referent of the field to the decrement buffer and adds the address of the field to the \emph{modified field buffer}.  Note that only this slow path, that records the field the first time it is written, is synchronized.  The \textjava{attemptToLog()} function will block if another thread is logging, returning only after the other thread has captured the to-be-overwritten value. Since objects are zeroed prior to allocation, they are naturally initialized with the unlogged bit/s cleared.  The barrier therefore does not log changes to new objects, cleanly implementing the \impdead optimization (\cref{sec:rc}, \cite{SBF:12}).

When \lxr processes the modified fields buffer, it generates an increment for referents of each field  and resets its unlogged bit. LXR uses entries in the \textjava{decbuf} for RC decrements and to create the SATB snapshot.  During RC pauses, LXR dereferences entries in the \textjava{modbuf} to find final referents, which receive RC increments.  The geometric mean  overhead of this write barrier 
compared to no write barrier on Immix is 1.6\%, with a worst case for \textsf{h2o} at 4.6\% (\cref{tab:analysis}).

\subsection{Parallelism}
\label{sec:parallelism}

\lxr is designed for scalability.  For mutator scalability, \lxr uses lock free block allocators to issue clean and recycled blocks to thread-local allocators with minimal contention, even at very high allocation rates.  For collector scalability, coalescing reference counting is naturally scalable. Reference counts are inherently local, so applying increments and decrements is embarrassingly parallel. We found partitioning very large reference arrays necessary for scalability of parallelized increment processing.
We use work stealing within the collector to maximize load balancing.  Only the concurrent SATB trace has significant potential for scalability issues~\cite{BP:10}.   Because it is a tertiary collection mechanism (following young object treatment and reference counting), this concern is not a major one and rarely encountered in practice. Most decrements are processed concurrently with limited  impact on { pause times, immediacy, or mutator performance.}



%% file: methodology.tex
\section{Methodology}
\label{sec:methodology}
\label{sec:method:hw}

We implement \lxr in the new MMTk~\cite{BCM:04a,BCM:04b,MMTk:22} on OpenJDK, fork jdk-11.0.11+6.\footnote{Commit 1007c9c1f46644d325bb4d4bd3a3d6dc718c713e, 2021/3/10.} We use G1, Shenandoah, and ZGC on the same fork. C4 is not available in OpenJDK. Because our implementation is missing class unloading, compressed pointers, and weak references (see \cref{sec:threats:cp}), we disable them in all systems. The implementation we evaluate here is publicly available~\cite{WZ:22}, and we expect it to become integrated and maintained as part of MMTk's public release~\cite{MMTk:22}. 
We use three hardware platforms described in~\cref{tab:machines}. The \zenthree uses a solid-state drive while the \zentwo and \coffeelake use hard disk drives. Unless otherwise stated, we report AMD \zenthree results.  We use a Intel Coffee Lake (\coffeelake) to assess the robustness of our findings to microarchitecture.   All machines run Ubuntu 18.04.  The remainder of this section describes \lxr configurations, benchmarks,  and measurement methodologies.

\begin{table}
  \caption{We use three contemporary stock processors.  We report results from the \zenthree by default.}
  \vspace*{-2ex}
  \input{tables/machines}
  \vspace*{-1.6ex}
    \label{tab:machines}
\end{table}

\subsubsection*{LXR Configuration}
\label{sec:configs}

Unless otherwise stated, we use the following configuration of LXR for performance measurements: a 4\,MB lock-free global block allocation buffer (\cref{sec:parallelism}), a two bit reference count, a 128\,MB survival threshold, no increment threshold (\cref{sec:alg-rctriggers}), a 5\% mature wastage threshold (\cref{sec:alg-satbtrig}),  and a single evacuation set (\cref{sec:alg-recmature}).
\subsubsection*{Benchmarks}
\label{sec:benchmarks}

We use 17 diverse up-to-date benchmarks from the Chopin development branch of the DaCapo benchmark suite~\cite{BGH+:06,DaCapo:21}.\footnote{Commit b00bfa96b6db296bdc6f57a57e56a9a34b2d2d89, 2022/04/09} It contains four latency-sensitive workloads that report request latency percentiles and throughput (total time).  We omit  \textsf{tradebeans} and \textsf{tradesoap} because they frequently fail on this version of JDK 11 unless the C2 compiler is disabled. 
ZGC fails on many benchmarks at 2$\times$ their nominal heap size because it requires a substantial minimum heap.  JDK 17 fixes this limitation. 

\cref{tab:stats} lists key benchmark characteristics. It presents the minimum G1 heap size, allocation characteristics, and object demographics. Allocation rates are relative to the mutator time of G1, excluding pause times. High allocation rates and high ratios of allocation to  minimum heap sizes  make most of these applications challenging workloads for collectors. Ten of 17 benchmarks allocate over a GB per second. The second-from-last column indicates the percentage of allocated bytes for objects larger than 16\,KB. The last column shows the percentage of bytes that survive a fixed 32\,MB nursery.  Nine benchmarks are highly generational with 0-5\% survival rates.  High survival rate is typically coupled with low ratios of allocation to heap size, with the exception of \textsf{xalan}. In contrast, \textsf{Batik} has a 51\% survival rate. 

\textsf{Cassandra, h2, lusearch,} and \textsf{tomcat} are request-based, latency-sensitive, and widely-used workloads, appropriate for evaluating latency optimized collectors. \textsf{Lusearch, h2,} and \textsf{tomcat} have high allocation rates, ranging from 1--9\,GB/s on the \zenthree.     \textsf{Cassandra} runs the YCSB workloads~\cite{CS+:10} on the Apache Cassandra NoSQL database management system (v. 3.11.10).  \textsf{H2} performs a TCP-C-like workload on the Apache Derby database (v. 10.14.2.0).  \textsf{Lusearch} performs half a million search queries on the Apache Lucene search framework  (v. 8.8.2).  It is highly parallel and has a small heap footprint.  \textsf{Tomcat} runs the sample web application workload provided with the Apache Tomcat web server (v. 9.0.45). 

\begin{table}
  \caption{Benchmark characteristics: minimum G1 heap size; total bytes allocated; ratio of allocation to minimum heap; allocation rate; mean object size; percentage of large object bytes to total bytes; and percentage of survivor bytes to total bytes for a 32\,MB nursery.
    }\vspace*{-2.6ex}

    \input{tables/bm-stats}
\vspace*{-.4ex}
    \label{tab:stats}
\end{table}

\subsubsection*{Throughput Measures}
\label{sec:method:throughput}

We report throughput as total time. We invoke each collector on each benchmark 20 times, perform five iterations, and report the time for the fifth iteration on an otherwise empty machine. This methodology gives the compiler and runtime time to warm up. We calculate the geometric mean and 95\% confidence intervals on 20 measurements.  Our measurements interleave collectors to minimize bias due to systemic interference.

\subsubsection*{Latency Measures}
\label{sec:method:latency}

The Chopin development version of DaCapo reports  \emph{simple} and \emph{metered} latency measurements.  Three significant factors impact request latency:
\begin {enumerate*}[i)]
\item the uninterrupted time to \emph{compute} the request,
\item the time consumed by \emph{interruptions} such as scheduling and garbage collection, and
\item the time consumed by \emph{request queuing}.
\end{enumerate*}
The simple measurement captures the first two, but not the third. DaCapo's metered measurement  captures all three by modeling request queuing, which is  standard in cloud-based services.  Requests arrive at servers at some \emph{metered} rate. The server enqueues these requests if there are insufficient resources to process them immediately. When an interruption arises (such as a garbage collection event), it delays both  active requests and any enqueued requests.  DaCapo models an unbounded queue and arbitrarily long delays, whereas some systems drop requests.  
We present DaCapo's metered 50, 90, 99, 99.9 and 99.99 percentile tail latencies in tabular form and the full data in latency response curves (e.g., Figure~\ref{fig:latency3}).  Because they are so sensitive to noise, we use 40 invocations for all of our latency measurements.

%% file: tables/machines.tex
{
    \relsize{-1}{
\sffamily

\setlength{\tabcolsep}{.4ex}

\newcommand{\mytoprule}{\cmidrule[\heavyrulewidth]{1-1}\cmidrule[\heavyrulewidth]{3-4}\cmidrule[\heavyrulewidth]{6-10}}
\newcommand{\mymidrule}{\cmidrule{1-1}\cmidrule{3-3}\cmidrule{5-7}}
\newcommand{\mybottomrule}{\mytoprule}
\begin{tabular}{l c@{\hspace{1ex}} r c@{\hspace{1ex}} r r r c@{\hspace{1ex}} r r}
\otoprule

\textbf{Name} && Model && Cores & Clock & LLC && \multicolumn{2}{c}{Memory}\\
\cmidrule(){1-1}\cmidrule(){3-3}\cmidrule(){5-7}\cmidrule(){9-10}
\zenthree && AMD 5950X && 16/32 & 3.4\,GHz & 64\,MB && 64\,GB & DDR4 3200MHz \\
\zentwo && AMD 3900X && 12/24 & 3.8\,GHz & 64\,MB && 64\,GB & DDR4 2133MHz \\
\coffeelake && Core i9-9900K && 8/16 & 3.6\,GHz & 16\,MB && 128\,GB & DDR4 3200MHz \\
\bottomrule

\end{tabular}
    }
}

%% file: tables/bm-stats.tex
{
\sffamily

\setlength{\tabcolsep}{0.7ex}

\begin{tabular}{l c@{\hspace{0.3ex}} r c@{\hspace{0.3ex}} r r r c@{\hspace{0.3ex}} r c@{\hspace{0.3ex}} r r}
    \\[-2ex]
    \otoprule
                       &  & \multicolumn{1}{c}{Heap} &  & \multicolumn{3}{c}{Allocation} &       & Obj     &  & \multicolumn{2}{c}{\%}                 \\

    \textbf{Benchmark} &  & \multicolumn{1}{c}{MB}   &  & GB                             & /heap & MB/s    &  & Size                   &  & Lrg & Srv. \\
    \cmidrule(){1-1}\cmidrule(){3-3}\cmidrule(){5-7}\cmidrule(){9-9}\cmidrule(){11-12}
    cassandra	&	&\,263	&	&5.6	&22	&\,596	&	&\,50	&	&0	&4	\\
    h2	&	&1\,191	&	&13.0	&11	&1\,534	&	&\,64	&	&0	&17	\\
    lusearch	&	&\,53	&	&31.2	&603	&9\,520	&	&\,97	&	&1	&1	\\
    tomcat	&	&\,71	&	&6.9	&100	&1\,440	&	&\,95	&	&21	&1	\\
    \cmidrule(){1-1}\cmidrule(){3-3}\cmidrule(){5-7}\cmidrule(){9-9}\cmidrule(){11-12}
    avrora	&	&\,7	&	&0.2	&28	&\,46	&	&\,45	&	&0	&5	\\
    batik	&	&1\,076	&	&0.5	&0	&\,257	&	&\,71	&	&10	&51	\\
    biojava	&	&\,191	&	&11.8	&63	&\,800	&	&\,37	&	&3	&2	\\
    eclipse	&	&\,534	&	&8.3	&16	&\,595	&	&\,100	&	&29	&17	\\
    fop	&	&\,73	&	&0.5	&7	&\,557	&	&\,58	&	&3	&10	\\
    graphchi	&	&\,255	&	&11.9	&48	&1\,117	&	&\,134	&	&3	&4	\\
    h2o	&	&3\,689	&	&11.8	&3	&3\,065	&	&\,168	&	&23	&14	\\
    jython	&	&\,325	&	&5.2	&16	&1\,038	&	&\,60	&	&4	&0	\\
    luindex	&	&\,41	&	&2.2	&54	&\,335	&	&\,288	&	&75	&3	\\
    pmd	&	&\,637	&	&7.0	&11	&3\,952	&	&\,46	&	&2	&14	\\
    sunflow	&	&\,87	&	&20.5	&241	&6\,267	&	&\,45	&	&0	&3	\\
    xalan	&	&\,43	&	&3.9	&92	&4\,265	&	&\,122	&	&41	&17	\\
    zxing	&	&\,153	&	&1.5	&10	&1\,750	&	&\,183	&	&50	&23	\\
    \bottomrule
\end{tabular}
}

%% file: results.tex
\section{Evaluation}
\label{sec:results}

We first compare \lxr to G1, Shenandoah, and ZGC with respect to request latency on the four latency-sensitive applications. We then present throughput and the sensitivity results.  
Because garbage collection is a time-space tradeoff, we explore sensitivity to heap size.   
Our results show that \lxr combines low latency with high throughput. Compared to the low latency collectors (Shenandoah, ZGC) and G1,
  \lxr matches or improves over their latency on all four benchmarks. {On throughput, \lxr improves over G1 by 4\% and Shenandoah by 43\% on our benchmarks.} 

\subsection{Request Latency}
\label{sec:res-latency}

\begin{figure*}  
  \captionof{table}{Median and tail request latencies in a 1.3$\times$ heap on the \zenthree with 95\% confidence intervals.  Best results in green.}
  \vspace*{-1em}
    \input{tables/latency-3}
    \label{tab:latency3}
    \centering
\end{figure*} 

\begin{figure}
    \newlength{\subcaptionsquish}
    \setlength{\subcaptionsquish}{-1.1ex}
    \begin{subfigure}[b]{0.495\columnwidth}
        \centering
        \includegraphics[width=\columnwidth]{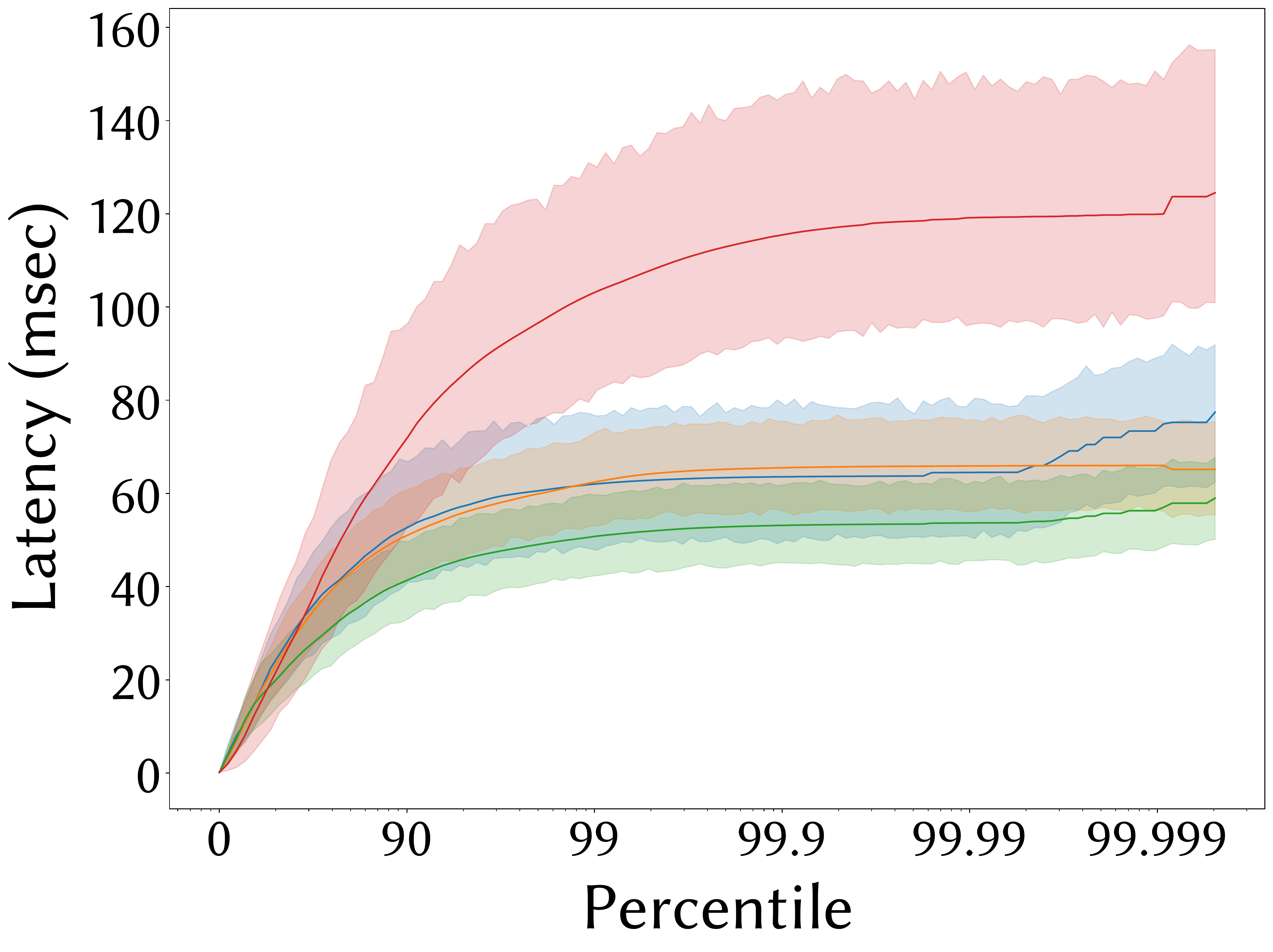}\vspace*{\subcaptionsquish}
        \caption{cassandra}
        \label{fig:latency-cassandra}
    \end{subfigure}
    \hfill
    \begin{subfigure}[b]{0.495\columnwidth}
        \centering
        \includegraphics[width=\columnwidth]{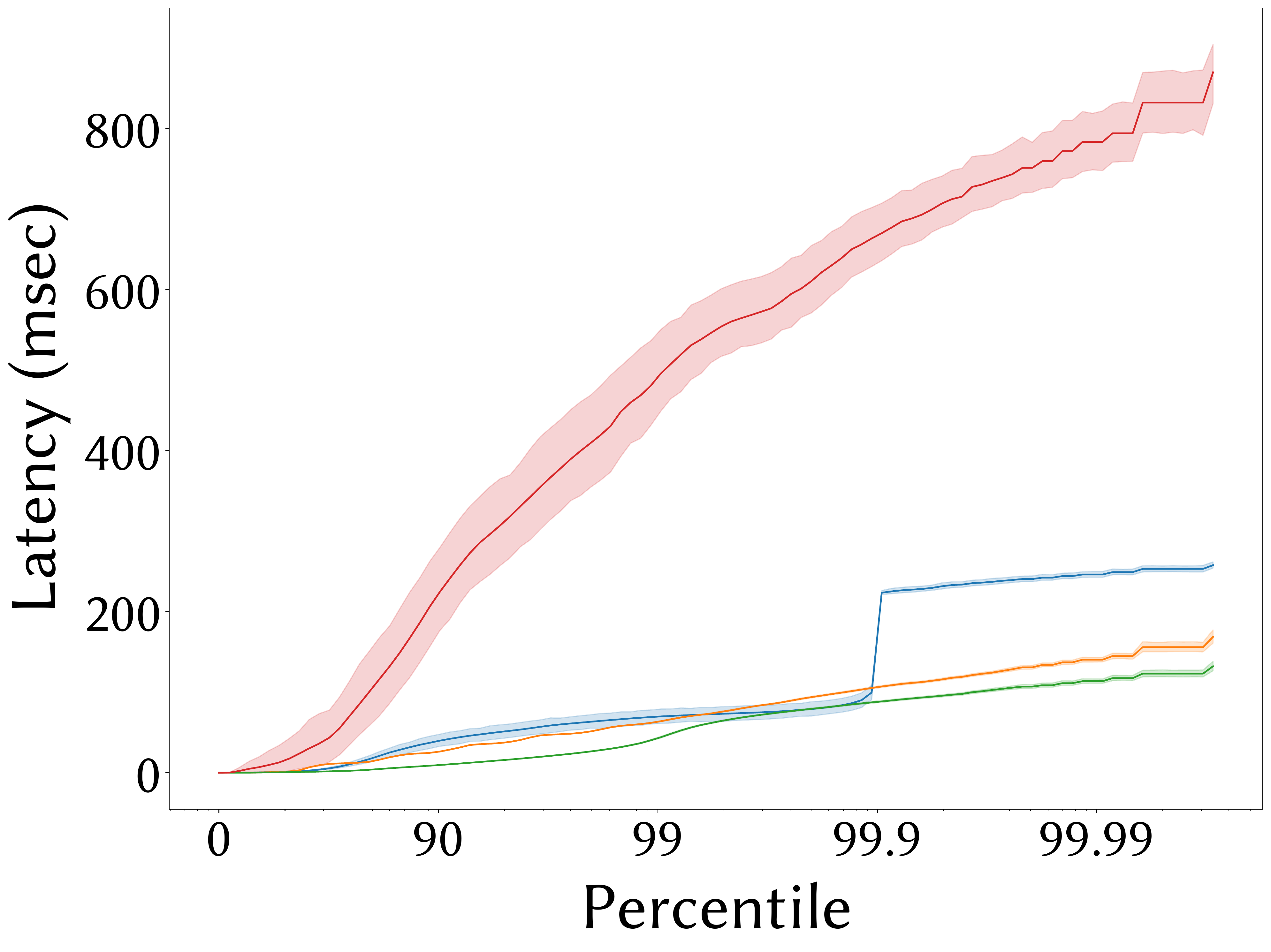}\vspace*{\subcaptionsquish}
        \caption{h2}
        \label{fig:latency-h2}
    \end{subfigure}
    \vskip\baselineskip\vspace{-0.5ex}
    \begin{subfigure}[b]{0.495\columnwidth}
        \centering
        \includegraphics[width=\columnwidth]{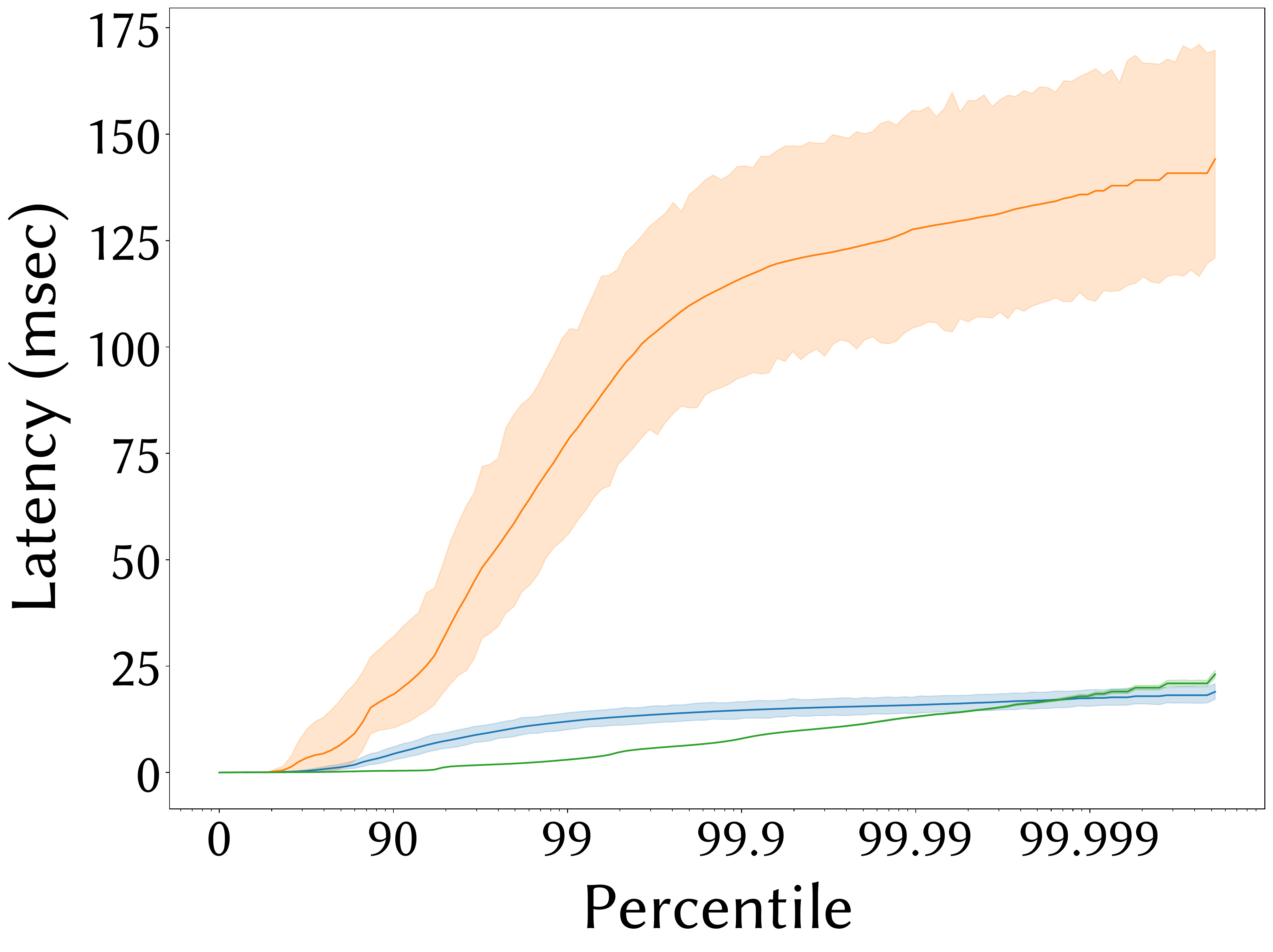}\vspace*{\subcaptionsquish}
        \caption{lusearch}
        \label{fig:latency-lusearch}
    \end{subfigure}
    \hfill
    \begin{subfigure}[b]{0.495\columnwidth}
        \centering
        \includegraphics[width=\columnwidth]{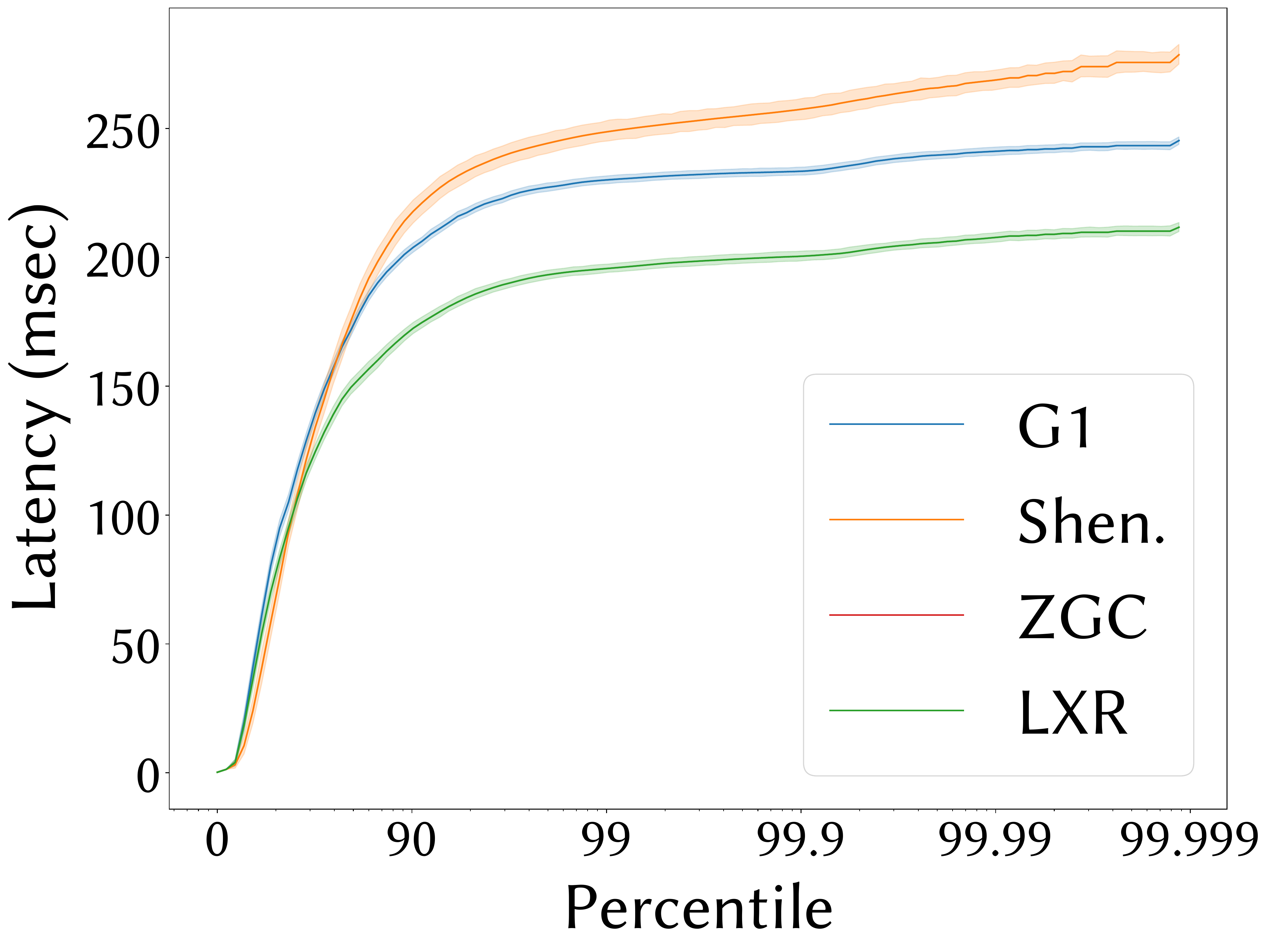}\vspace*{\subcaptionsquish}
        \caption{tomcat}
        \label{fig:latency-tomcat}
    \end{subfigure}
    \captionof{figure}{Request latency response curves on the \zenthree with a  1.3$\times$ heap.  Each y-axis shows  latency in milliseconds and the x-axis shows the percentile of requests experiencing that latency.   `90' on the x-axis represents the latency observed by the 90th percentile of requests.  Each solid line represents the average from 40 invocations and the shaded area shows its 95-percentile variance. \lxr improves over the other collectors at nearly every point of every curve. 
    }\vspace*{-1ex}
    \label{fig:latency3}
\end{figure} 

We evaluate G1, \lxr, Shenandoah, and ZGC on the four latency-sensitive workloads described in~\cref{sec:method:latency}.  \cref{tab:latency3} and \cref{fig:latency3} report results for a tight 1.3$\times$ heap.   
This version of ZGC cannot run in small heaps so produced results only for \textsf{cassandra} and \textsf{h2}.
\cref{tab:latency3} shows the median, 90, 99, 99.9 and 99.99 percentile request latencies and the 95\% confidence intervals, which are expressed as a fraction:   \cival{0.500}{0.500} means the 95\% confidence interval extends 50\% over the reported result in each direction. \cref{fig:latency3}(a) graphs the mean at  all request percentiles with their 95\% confidence intervals shaded.

\lxr performs better or as well as the other collectors at essentially all percentiles.
Recall the three components of a latency measure (\cref{sec:method:latency}) and that variability in the computation of the requests may dominate.
For \textsf{cassandra}, noise, regardless of the collector, reduces the differences between \lxr, Shenandoah,  and  G1, as shown in \cref{fig:latency3}(a).

  For \textsf{h2}, \lxr outperforms both latency-targeting collectors at all percentiles.  G1 performs similar to Shenandoah until the 99.9th percentile, when its latency roughly doubles, but it substantially outperforms ZGC at every percentile. The performance  on \textsf{h2} with a minimum heap size of 1.2\,GB demonstrates that \lxr  performs well in large heaps  running traditional workloads such as TPC-C, as well as performing well on the other workloads with smaller minimum heap sizes (\cref{tab:stats}).

For \textsf{lusearch}, \lxr and G1 are similar, but  Shenandoah performs poorly, as it currently lacks an effective mechanism for young objects and very high allocation rates.
\textsf{Lusearch} has a small heap size, a very high allocation rate, and a very low survival rate in contrast to \textsf{h2}.  \lxr's excellent performance on both \textsf{h2} and \textsf{lusearch} shows its design at its best, delivering excellent results on such different workloads.  On the other hand, while G1 does well on \textsf{lusearch}, it does not do well in the tail on \textsf{h2}, and  Shenandoah does well on \textsf{h2}, but struggles with \textsf{lusearch}'s high allocation rate.

For \textsf{tomcat}, \lxr delivers large improvements at the 90th percentile and higher. The narrow confidence intervals and consistency in the result among the three collectors suggests that  request latency is, like \textsf{cassandra}, dominated by the intrinsic computational cost of each of benchmark's request more so than interruptions or request queuing.   The better responsiveness of \lxr is nonetheless visible in \cref{fig:latency3}(d).

These four applications exhibit diverse behaviors, such as high and low survival rates. These differences require the range of optimizations in \lxr.   \cref{tab:stats} shows the high survival rate of \textsf{h2}  at 17\%, whereas the others have low survival rates: \textsf{cassandra} at 4\%, \textsf{lusearch} at  1\%, and \textsf{tomcat} at 1\%. \lxr's RC and \impdead optimization are extremely effective on the three generational workloads. Mature space optimizations are however critical for all benchmarks,  even \textsf{lusearch} and \textsf{tomcat}, because their ratio of allocation to minimum heap size is high (\cref{tab:stats}).

Mature space RC is an effective optimization for these benchmarks. \cref{tab:analysis} shows that \lxr's  mature object RC (\textsf{Old}) reclaims many mature objects.  Mature object RC reclaims
60\% of mature objects for \textsf{cassandra},
48\%  for \textsf{h2},  
40\% for \textsf{lusearch},
and 71\% for \textsf{tomcat}. 
The SATB trace reclaims the other mature objects.
We note that \lxr's SATB trigger successfully tunes itself to the effectiveness of the RC mature reclamation behavior (see \cref{tab:analysis} under GC Pauses with SATB\%), running less frequently (14\% of RC pauses) on \textsf{tomcat} versus 20-21\% for the others, since \lxr's mature RC reclaims most mature objects for \textsf{tomcat}.

\subsubsection*{Heap Size Sensitivity}
The above analysis uses a tight, 1.3$\times$ heap to highlight the time-space tradeoff inherent to garbage collection and how that affects application latency.  \cref{tab:heapsize} shows  the heap size sensitivity of request latency and performance with a  2$\times$ heap and a  generous 6$\times$ heap. It presents the geometric mean latency for the four latency benchmarks and time for all benchmarks. For latency, \lxr outperforms G1 and Shenandoah at each size, and at 1.3$\times$ and 6$\times$ \lxr is substantially better.
On throughput,  \lxr is better than G1 in tighter heaps and substantially better than Shenandoah. These results illustrate that \lxr successfully pairs low latency and high throughput on a range of heap sizes.

\begin{table}
  \caption{{Geometric mean of 99.99\% latency for \textsf{cassandra}, \textsf{h2}, \textsf{lusearch}, and \textsf{tomcat} and throughput for \emph{all benchmarks}. Results relative to G1 for each collector on three heap sizes.}}
  \vspace*{-1em}
\input{tables/heapsize}
  \label{tab:heapsize}
 \vspace*{-1em}
\end{table}

\subsection{Throughput}
\label{sec:res-throughput}

\begin{table}
    \caption{Benchmark throughput with a 2$\times$ heap.  We report G1 time  in milliseconds and relative performance of \lxr, Shenandoah, and ZGC, normalized to G1. Best results for each collector are shown in green, worst are shown in orange.}
    \vspace*{-1em}
  \input{tables/time-3}
 \vspace*{-1em}
    \label{tab:time3}
\end{table}

\cref{tab:time3} reports throughput using a moderate heap, {2$\times$} the minimum heap for G1. Column two reports running time for G1 in milliseconds. 
The next three columns show relative performance for \lxr, Shenandoah, and ZGC. 
The confidence intervals (unshown) are less than 1\% for most systems, with the highest value on \textsf{sunflow} at 4\%.
\cref{tab:heapsize} shows that the throughput results discussed below hold up on a wide range of heap sizes.
ZGC cannot run some workloads because the heap size is too small. 

  \lxr's average speedup is 4\% over G1 and 43\% over Shenandoah.   Among the latency-sensitive workloads, 
  \lxr  offers a 4\% and 9\% improvement over G1 on  \textsf{cassandra} and \textsf{lusearch}. On \textsf{h2}, G1 has the best performance, whereas \lxr is slower by 10\%, but Shenandoah and ZGC are substantially slower at 25\% and 76\%, respectively.   On \textsf{lusearch}, Shenandoah is 6.6$\times$ slower than G1, as it still struggles to keep up with the extremely fast rate at which Lucene generates garbage. 
 Shenandoah is within 3\% of G1 on five benchmarks, but otherwise it pays for its short pauses with more pauses (measured but unshown) and more expensive barriers, degrading mutator and total performance.
 Shenandoah's best result is on \textsf{eclipse} which has  a high 17\% survival rate.


\begin{table*}
    \caption{Breaking down \lxr.  The first column indicates execution time. The next three columns show the impact of \textbf{concurrency optimizations} relative to \lxr by turning off concurrency for the trace (-SATB) and for lazy decrements (-LD), and for both (STW).  The next five columns show stats for \textbf{GC Pauses}: pauses/second, 50th and 95th percentile pauses, fraction due to SATB, and percentage that occur before lazy decrements complete.   The next two capture key \textbf{barrier} statistics: increments per ms and the overall field barrier overhead. The last five columns show key \textbf{reclamation} statistics: percentage of reclamation due to young RC objects, old RC objects, SATB, percentage of old objects with a stuck RCs, {and the ratio of evacuated young volume over the reclaimed young clean blocks (YC).}}
    \vspace*{-.75em}
  \input{tables/lxr-analysis}
    \label{tab:analysis}
\end{table*}

By contrast, \lxr's young object optimizations  give it an advantage on \textsf{lusearch} and other low survival rate benchmarks  (\cref{sec:alg-recyoung}, \cite{SBF:12}). \lxr offers substantial improvements over G1 on \textsf{avrora}, \textsf{sunflow}, and \textsf{xalan}: 12\%, 22\%, and 36\% respectively, and even higher improvements over Shenandoah. These benchmarks are less generational, with survival rates of 5\%, 3\%, and 17\%, demonstrating \lxr's ability to adapt to diverse application behaviors. On these benchmarks \lxr is working as intended,  modulating the  SATB tracing to about 20\% of collections, where the SATB trace reclaims most of the mature dead objects (\cref{tab:analysis}).


Examining \lxr's excellent performance on \textsf{avrora} in detail,  we find that G1 and Shenandoah are spending significantly amounts of time in concurrent collection processing a large, long-lived linked list.  
This list creates a bottleneck for the tracing collectors~\cite{BP:10}. The live list has much less effect on \lxr because it only occasionally performs backup SATB tracing. \textsf{Avrora} demonstrates that the worst case limitations when relying only on tracing and copying are not just theoretical, but are encountered in practice.  On the benchmarks where \lxr performs the best (\textsf{avrora}, \textsf{sunflow}, and \textsf{xalan}),  \lxr has substantially fewer RC pauses,  retired instructions, and cache misses (measured but unshown) than G1 and Shenandoah. \lxr's improvements over Shenandoah highlight the CPU and memory penalties  of region-based concurrent copying.


\subsection{Analysis of Overheads}
\label{sec:res-overheads}

This section explores the features and overheads of \lxr. \cref{tab:analysis} presents results for \lxr in a 2$\times$ heap. 
The first column indicates the running time for each benchmark. 

\subsubsection*{Concurrency}

The next three columns of \cref{tab:analysis} show the performance impact of turning off concurrency features by i) tracing (-SATB) in the RC pause, ii) decrementing  (-LD) in the RC pause, and iii) both (STW), yielding a stop-the-world collector. It shows  ratios with respect to the default \lxr. 

Tracing concurrently or in the pause has the same total performance on these workloads. \textsf{H2} degrades the most with stop-the-world tracing, adding {4}\% to execution time, whereas \textsf{xalan} speeds up by {18}\%.   Decrementing in the pause slows the benchmarks down by {3}\% on average, and slows down \textsf{h2} by {26}\%! These results show that increases in pause time work versus concurrency with the application often make little impact on total time.  \textsf{Xalan} is an outlier. It speeds up when the concurrency features are turned off. This result shows that concurrency features sometimes interfere with application threads and doing them when the mutators are already paused can be efficient with respect to total time.

With both SATB tracing and decrements in the pause (a fully stop-the-world collector), the benchmarks run {3}\% slower on average. \textsf{H2} has the worst  slow down  at {31}\% and \textsf{xalan} improves by {9}\%.  \lxr outperforms STW, an approximation of RC-Immix~\cite{SBYM:13}.  However, STW lacks the carefully tuned whole heap defragmentation heuristics RC-Immix used. 
  Summarizing,  the concurrency features usually do not penalize throughput and are essential to low pause times.

  \begin{figure*}[h]
    \setlength{\subcaptionsquish}{-1.1ex}
    \begin{subfigure}[t]{\columnwidth}
        \centering
        \includegraphics[width=\columnwidth]{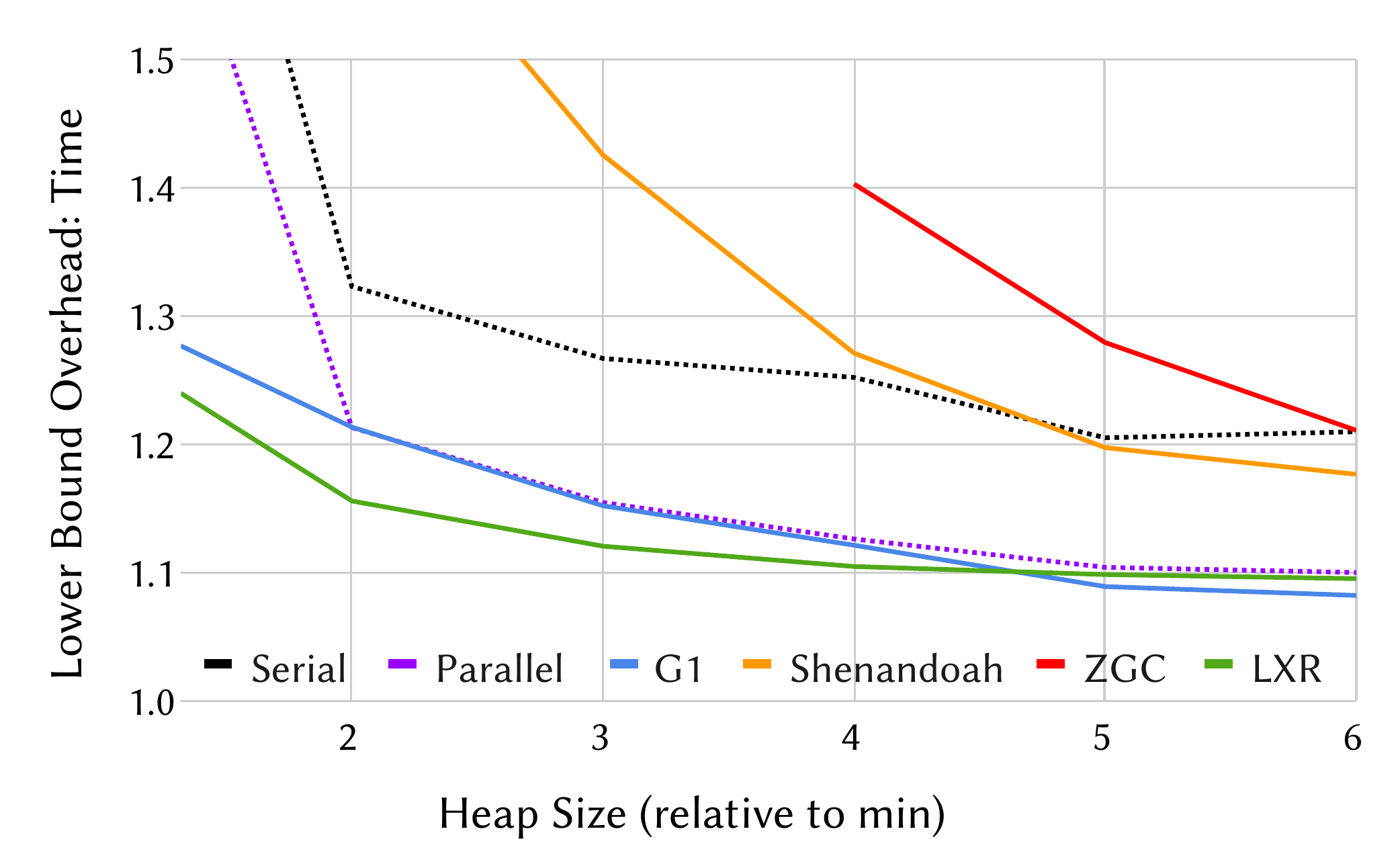}\vspace*{\subcaptionsquish}
        \caption{Overhead: wall-clock time.  In all but the largest heap sizes, LXR outperforms all collectors.}
        \label{fig:lbo-time}
    \end{subfigure}\hspace{2em}
    \begin{subfigure}[t]{\columnwidth}
        \centering
        \includegraphics[width=\columnwidth]{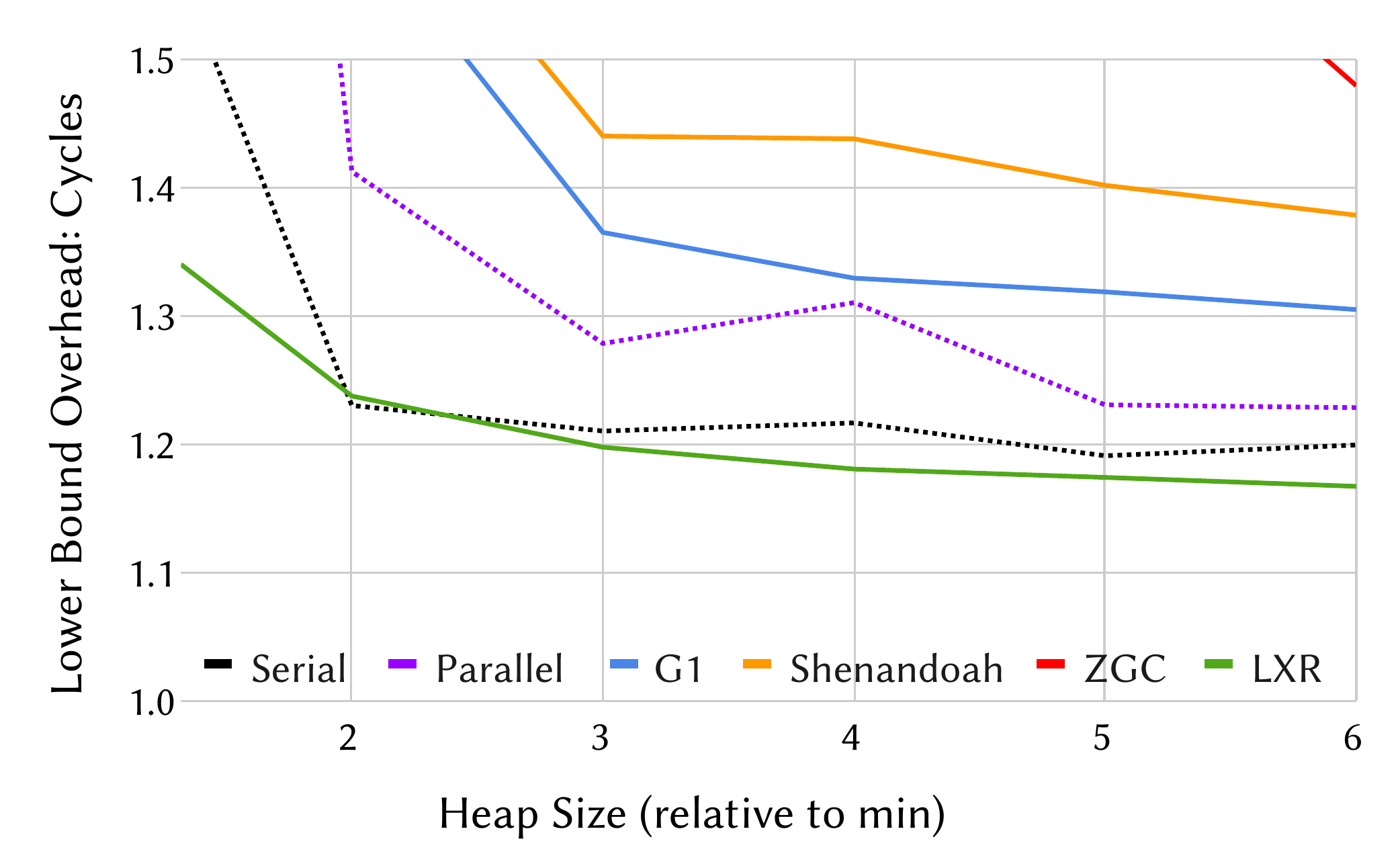}\vspace*{\subcaptionsquish}
        \caption{Overhead: total cycles.  This reflects \emph{all} work done by or on behalf of the collector, including concurrent threads. LXR's algorithmic advantage is clear at every heap size.  
        }
        \label{fig:lbo-cycles}
    \end{subfigure}
    \captionof{figure}{Lower bounds on collector overheads averaged over all benchmarks, relative to a notionally ideal collector following the LBO methodology~\cite{CBBM:22}. Each line represents the average overhead over all benchmarks for 40 invocations. OpenJDK's two stop-the-world collectors, Serial and Parallel, are shown with dashed lines.
    }\vspace*{-1ex}
    \label{fig:lbo}
  \end{figure*} 

\subsubsection*{GC Pauses} 
{The next five columns characterize collector pauses, showing their frequency (/s), the 50th and 95th percentile pause time in ms, the fraction of pauses that  start an SATB trace, and the fraction of pauses that occur before lazy decrements complete.  The frequency of pauses varies by two orders of magnitude, with \textsf{lusearch} and \textsf{xalan} having the highest rate, at about one every 10\,ms.  The median and 95th percentiles of pause durations are {5.0} and {7.5}\,ms respectively, on average.  \textsf{H2}, \textsf{batik},  and \textsf{pmd} have the longest pauses.   
The number of collections that trigger an SATB 
is 21\% on average and as high as 50\% for \textsf{batik}.  It is rare for lazy decrement processing to run into the next pause.  For most benchmarks, it never happens, but in the worst case, \textsf{xalan}, it reaches  22\%.}

\subsubsection*{Barriers}  The next two columns of \cref{tab:analysis} show the number of write barriers per ms and the  field barrier overhead. 
The rate of increments per millisecond directly corresponds to the barrier slow path take rate (\cref{fig:barrier}, line 3).  {This rate varies by more than two orders of magnitude from just 70 for \textsf{h2o} up to 9\,296 for \textsf{pmd}. 
  The total overhead of the field barrier is measured relative to no write barrier using full heap Immix.   On average the field barrier adds 1.6\% overhead compared to no barrier, ranging from no measurable overhead for  \textsf{avrora} and a still modest 4.6\% for \textsf{h2o}.

\subsubsection*{Reclamation}  The last four columns of \cref{tab:analysis} show the percentage of objects reclaimed via the \impdead RC optimization (Young), mature RC (Old), and by the SATB trace; followed by the percentage of objects with stuck reference counts {and the ratio of young object bytes evacuated to blocks freed (YC).  At this 2$\times$ heap size, the vast majority of objects are reclaimed via the \impdead optimization. In the case of \textsf{biojava}, \textsf{h2o}, and \textsf{jython}, it reclaims 100\%. \textsf{Batik} has the highest number of objects reclaimed by the SATB trace, at 54.5\%, with \textsf{pmd} at 6.9\%.  The second last column shows that very few objects have stuck reference counts, at most 6.6\%, for \textsf{batik}.  The last column shows the number of bytes evacuated from young blocks relative to blocks reclaimed.   The outlier is \textsf{batik}, which has a very high survival rate, so has to perform a lot of copying to yield free blocks.  However, the geometric mean is just 1.1\%,  illustrating  that \lxr  reclaims most space without performing copying.

\subsection{Sensitivity Analysis}
\label{sec:res-sensitivity}

This section summarizes the sensitivity of \lxr to
\begin {enumerate*}[i)]
\item block size,   
\item number of reference counting bits, 
\item size of the lock free block allocator, and %
\item {architecture}. 
\end{enumerate*}  

{\lxr uses a block size of 32\,KB, and we found that halving it improved performance by 0.6\% on average but one benchmark (\textsf{pmd}) did not run to completion, while doubling introduced a 3.9\% overhead and \textsf{h2o} ran out of memory.}
{\lxr uses 2 reference counting bits per object.  We evaluated 4 and 8 bits. Since address arithmetic 
  maps objects to their counts, we considered only power-of-two bit counts.  \lxr is sensitive to this choice, with a 2.9\% penalty for 4 bits and a 3.4\% penalty for 8 bits. This result is notable, demonstrating that just two bits is quite effective for reference counting these workloads. 
%

{\lxr uses a 32 entry lock-free free block buffer} by default (\cref{sec:parallelism}). We evaluated both 64 and 128 entry buffers.  The differences were small, just 1.1\% and 1.3\% slower respectively.   Notably, \textsf{lusearch}, which is the fastest allocating workload (\cref{tab:stats}) and motivated us to introduce this structure, saw a 12\% slowdown for a 128-entry buffer.

{
  We also explored sensitivity to architecture on three machines, each with a different core count,  microarchitecture, and  clock speed.  At a 2$\times$ heap, \lxr's throughput relative to G1 was robust to architecture, showing 4.2\%, 3.4\%, and 2.2\% geometric mean performance improvements over G1 on the \zenthree, \zentwo, and \coffeelake respectively.   Shenandoah showed a similar trend but much greater sensitivity, reporting 37.3\%, 29.4\%, and 19.6\% slowdowns respectively.   We observed similar patterns at the other heap sizes.
}

\subsection{LBO Analysis}
\label{sec:res-lbo}

\cref{fig:lbo} presents a lower bound overhead (LBO) analysis of LXR, G1, Shenandoah, ZGC, and OpenJDK's two stop-the-world collectors, Serial and Parallel.  \citeauthor{CBBM:22} introduced the LBO methodology as a way of placing an emperical lower bound on the overheads introduced by garbage collectors~\cite{CBBM:22}.  

\citeauthor{CBBM:22}'s approach is to establish a baseline that is an approximation to an ideal (zero cost) garbage collector, and then compare total execution costs against that baseline.  The LBO methodology can be applied with respect to any performance metric.  Here we evaluate wall clock time and CPU cycles.  Generally speaking, some sources of GC overhead are hard to separate such as allocator costs, write barriers, and concurrent collection. On the other hand, costs due to stop-the-world pauses are easy to measure and account for.  The LBO methodology utilizes this insight to establish the baseline for each metric being evaluated.   It does this by evaluating each benchmark against a suite of collectors, subtracting the easy to measure stop-the-world GC costs in each case, and using the lowest cost execution as the baseline for that combination of benchmark and metric.  In practice, the collectors that yield the lowest cost tend to be simple ones such as semi-space, which have high stop-the-world overheads but have few other overheads and deliver good mutator locality.  We included semi-space in our LBO analysis in addition to each of the other collectors we evaluate. We found that the best baseline was most often produced by Parallel for wall clock time (7/16 benchmarks), and by semi-space for CPU cycles (7/16). Since by definition the baseline is \emph{at least} as expensive as the true ideal, the derived overhead is a \emph{lower bound} on the true overhead of collection.  LBO reports the ratio between the measured collector and the baseline.   Thus, a collector with an LBO of 1.10 has an overhead of at least 10\% with respect to the ideal collector for the metric being evaluated.  We evaluate overhead at a range of heap sizes to expose the time-space tradeoff the collectors are making.

\cref{fig:lbo}(a) shows wall-clock time overhead for each of the collectors, revealing that LXR has the lowest overhead in all but the very largest of heaps.   The graph shows for example that the wall clock time for LXR in a 2$\times$ heap is 16\% longer than an approximation to the ideal collector, while G1 has a 21\% overhead and Shendandoah has an overhead of 66\% (not shown).  (ZGC could not run all benchmarks in smaller heap sizes, so only has data points for large heaps.)  Although the LBO analysis uses a different methodology, the findings are consistent with results in \cref{tab:time3}.

\cref{fig:lbo}(b) shows LBO with respect to total CPU cycles consumed.  The cycle count is integrated across all cores, and thus captures the total cost of the garbage collector and application, including work performed by concurrent garbage collection threads and by the mutator on behalf of the collector.  The results are striking.  LXR incurs substantially lower overhead than the other concurrent collectors at every heap size.  This result crisply highlights the fundamental algorithmic efficiency of LXR compared to existing collectors.   The other collectors spend considerable resources performing concurrent tracing and concurrent copying, neither of which are exposed by the wall clock time alone.   The algorithmic advantage of LXR is such that it has lower total overhead than even the most efficient throughput-oriented stop-the-world collectors, Serial and Parallel, in all but one data point.

%% file: tables/latency-3.tex
{
\sffamily
\setlength{\tabcolsep}{.55ex}

\newcommand{\mytoprule}{\cmidrule[\heavyrulewidth]{1-1}\cmidrule[\heavyrulewidth]{3-4}\cmidrule[\heavyrulewidth]{6-10}}
\newcommand{\mymidrule}{\cmidrule{1-1}\cmidrule{3-6}\cmidrule{8-11}\cmidrule{13-16}\cmidrule{18-21}}
\newcommand{\mybottomrule}{\mytoprule}
\newcommand{\pctl}[1]{\textcolor{gray}{\relsize{-1}{#1}}}


\begin{tabular}{l c@{\hspace{1ex}} r r r r c@{\hspace{4ex}} r r r r c@{\hspace{1ex}} r r r r c@{\hspace{1ex}} r r r r }
\otoprule
   && \multicolumn{4}{c}{\textbf{G1}}&&\multicolumn{4}{c}{\textbf{\lxr}}&&\multicolumn{4}{c}{\textbf{Shenandoah}}&&\multicolumn{4}{c}{\textbf{ZGC}}\\ 
   \textbf{Benchmark}&&\pctl{50}&\pctl{99}&\pctl{99.9}&\pctl{99.99}&&\pctl{50}&\pctl{99}&\pctl{99.9}&\pctl{99.99}&&\pctl{50}&\pctl{99}&\pctl{99.9}&\pctl{99.99}&&\pctl{50}&\pctl{99}&\pctl{99.9}&\pctl{99.99}\\
\mymidrule

cassandra	&			&24.2						&62.0			&63.6			&64.5		&				&\bst{19.9}						&\bst{50.8}			&\bst{53.1}			&\bst{53.7}		&				&23.2						&62.4			&65.5			&65.9		&				&21.7						&103.1			&115.5			&119.133			\vspace{-1ex}\\
	&			&\cival{0.332}{0.332}						&\cival{0.244}{0.244}			&\cival{0.237}{0.237}			&\cival{0.749}{0.231}		&				&\cival{0.342}{0.342}						&\cival{0.180}{0.180}			&\cival{0.172}{0.172}			&\cival{0.755}{0.170}		&				&\cival{0.286}{0.286}						&\cival{0.161}{0.161}			&\cival{0.156}{0.156}			&\cival{0.749}{0.155}		&				&\cival{0.571}{0.571}						&\cival{0.253}{0.253}			&\cival{0.235}{0.235}			&\cival{0.922}{0.230}			\\
h2	&			&1.1						&69.7			&104.6			&246.1		&				&\bst{0.6}						&\bst{42.9}			&\bst{87.7}			&\bst{113.6}		&				&1.3						&63.3			&105.9			&140.3		&				&16.3						&500.0			&677.1			&813.2			\vspace{-1ex}\\
	&			&\cival{0.112}{0.112}						&\cival{0.134}{0.134}			&\cival{0.089}{0.089}			&\cival{0.996}{0.015}		&				&\cival{0.009}{0.009}						&\cival{0.024}{0.024}			&\cival{0.013}{0.013}			&\cival{0.995}{0.026}		&				&\cival{0.014}{0.014}						&\cival{0.009}{0.009}			&\cival{0.012}{0.012}			&\cival{0.991}{0.022}		&				&\cival{1.481}{1.481}						&\cival{0.127}{0.127}			&\cival{0.078}{0.078}			&\cival{1.010}{0.073}			\\
lusearch	&			&\bst{0.1}						&12.0			&14.6			&15.9		&				&\bst{0.1}						&\bst{3.0}			&\bst{8.0}			&\bst{13.1}		&				&0.3						&78.0			&116.1			&127.8		&				& 						& 			& 			& 			\vspace{-1ex}\\
	&			&\cival{0.458}{0.458}						&\cival{0.170}{0.170}			&\cival{0.142}{0.142}			&\cival{0.997}{0.129}		&				&\cival{0.003}{0.003}						&\cival{0.020}{0.020}			&\cival{0.010}{0.010}			&\cival{0.996}{0.010}		&				&\cival{1.673}{1.673}						&\cival{0.313}{0.313}			&\cival{0.226}{0.226}			&\cival{1.001}{0.205}		&				& \\
tomcat	&			&89.3						&230.1			&233.4			&241.1		&				&77.8						&\bst{195.7}			&\bst{200.4}			&\bst{207.6}		&				&\bst{68.2}						&248.7			&257.5			&268.7		&				& 						& 			& 			& 			\vspace{-1ex}\\
	&			&\cival{0.040}{0.040}						&\cival{0.007}{0.007}			&\cival{0.007}{0.007}			&\cival{0.644}{0.007}		&				&\cival{0.045}{0.045}						&\cival{0.009}{0.009}			&\cival{0.009}{0.009}			&\cival{0.642}{0.009}		&				&\cival{0.088}{0.088}						&\cival{0.017}{0.017}			&\cival{0.016}{0.016}			&\cival{0.768}{0.015}		&				& \\

	\bottomrule
\end{tabular}
}

%% file: tables/heapsize.tex
{
\sffamily

\setlength{\tabcolsep}{0.6ex}

\begin{tabular}{r c@{\hspace{0.3ex}} r rr r c c@{\hspace{0.3ex}} r rr r c}
    \\[-2ex]
    \otoprule
    &  & \multicolumn{5}{c}{99.99\% Latency/G1} & \hspace*{1em} & \multicolumn{5}{c}{Time/G1}\\
    \multicolumn{1}{c}{Heap} &  & \multicolumn{1}{c}{G1} & &\multicolumn{1}{c}{\lxr} & \multicolumn{1}{c}{Shen} & \multicolumn{1}{c}{ZGC} & & \multicolumn{1}{c}{G1} & & \multicolumn{1}{c}{\lxr} & \multicolumn{1}{c}{Shen} & \multicolumn{1}{c}{ZGC} \\

            \cmidrule(){1-1}\cmidrule(){3-3}\cmidrule(){5-7}\cmidrule(){9-9}\cmidrule(){11-13}

    1.3$\times$ &&  1.00 && \bst{0.72} & 1.51 & -- &&1.00 && \bst{0.97} &1.77&--\\
    2$\times$  &   & 1.00 && \bst{0.92} & 2.54 & -- & &1.00 && \bst{0.96} & 1.37 & --\\
    6$\times$  &  & 1.00 && \bst{0.85} & 1.41 & 1.44& &\bst{1.00} && 1.01 & 1.09 & 1.26\\
    \bottomrule
\end{tabular}
}

%% file: tables/time-3.tex
{
  \setlength{\tabcolsep}{.35ex}
\sffamily
\newcommand{\mytoprule}{\cmidrule[\heavyrulewidth]{1-1}\cmidrule[\heavyrulewidth]{3-4}\cmidrule[\heavyrulewidth]{6-10}}
\newcommand{\mymidrule}{\cmidrule{1-1}\cmidrule{3-3}\cmidrule{5-7}}
\newcommand{\mybottomrule}{\mytoprule}
\begin{tabular}{l c@{\hspace{2ex}} r c@{\hspace{2ex}} r r r }
\\[-2ex]
\otoprule
  \textbf{Benchmark} && \multicolumn{1}{r}{\textbf{G1}}&&\multicolumn{1}{c}{\textbf{\lxr}}&\multicolumn{1}{c}{\textbf{Shen.}}&\multicolumn{1}{c}{\textbf{ZGC}}\\ 
\mymidrule                     
cassandra	&	&9\,718	&	&0.961	&1.015	&0.990	\\
h2	&	&8\,902	&	&\wst{1.101}	&1.246	&1.764	\\
lusearch	&	&3\,810	&	&0.909	&\wst{6.596}	&	\\
tomcat	&	&5\,082	&	&1.024	&1.220	&	\\
\mymidrule   
avrora	&	&4\,336	&	&0.881	&1.014	&	\\
batik	&	&1\,852	&	&1.015	&1.031	&\bst{0.989}	\\
biojava	&	&15\,215	&	&1.032	&1.107	&2.464	\\
eclipse	&	&14\,596	&	&1.019	&\bst{0.992}	&1.046	\\
fop	&	&\,934	&	&0.999	&1.159	&	\\
graphchi	&	&11\,032	&	&0.991	&1.079	&1.139	\\
h2o	&	&3\,972	&	&1.099	&1.002	&0.999	\\
jython	&	&5\,244	&	&1.018	&1.814	&\wst{2.793}	\\
luindex	&	&6\,638	&	&0.995	&1.071	&	\\
pmd	&	&2\,088	&	&0.992	&1.084	&1.124	\\
sunflow	&	&3\,717	&	&0.779	&1.450	&	\\
xalan	&	&1\,430	&	&\bst{0.643}	&4.784	&	\\
zxing	&	&\,871	&	&0.952	&1.030	&	\\\mymidrule
\multicolumn{1}{r}{\textbf{geomean}}	&	&	&	&\emph{0.958}	&\emph{1.373}	&	\\\bottomrule
\end{tabular}
}

%% file: tables/lxr-analysis.tex
{
\sffamily

\setlength{\tabcolsep}{.38ex}

\newcommand{\mytoprule}{\cmidrule[\heavyrulewidth]{1-1}\cmidrule[\heavyrulewidth]{3-4}\cmidrule[\heavyrulewidth]{6-11}}
\newcommand{\mymidrule}{\cmidrule(){1-1}\cmidrule(){3-3}\cmidrule(){5-7}\cmidrule(){9-13}\cmidrule(){15-16}\cmidrule(){18-20}\cmidrule(){22-22}\cmidrule(){24-24}}
\newcommand{\mybottomrule}{\mytoprule}
\begin{tabular}{l c@{\hspace{1ex}} r c@{\hspace{1ex}}  rrr c@{\hspace{1ex}} rrrrr c@{\hspace{1ex}} rr c@{\hspace{1ex}} rrr c@{\hspace{0.5ex}} r c@{\hspace{0.5ex}} r}
\\[-2ex]
\otoprule

 &&\multicolumn{1}{c}{time}&&  \multicolumn{3}{c}{Concurrency} &&  \multicolumn{5}{c}{GC Pauses} && \multicolumn{2}{c}{Barriers}&& \multicolumn{7}{c}{Reclamation (\%)}\\
 \textbf{} && \multicolumn{1}{c}{ms} && \multicolumn{1}{c}{-SATB} &\multicolumn{1}{c}{-LD} &\multicolumn{1}{c}{STW} && \multicolumn{1}{c}{/s} &\multicolumn{1}{c}{50\%\,ms}&\multicolumn{1}{c}{95\%\,ms} & \multicolumn{1}{c}{SATB\%}&\multicolumn{1}{c}{!Lazy\%}&&\multicolumn{1}{c}{Inc/ms} &\multicolumn{1}{c}{o/h}  &&\multicolumn{1}{c}{Young}&\multicolumn{1}{c}{Old}&\multicolumn{1}{c}{SATB}&&\multicolumn{1}{c}{Stuck}&&\multicolumn{1}{c}{YC}\\
 \mymidrule
cassandra	&	&9\,335	&	&1.01	&1.05	&1.05	&	&2.0	&4.6	&7.4	&21	&0	&	&\,399	&0.991	&	&98.5	&0.9	&0.6	&	&0.0	&	&1.2	\\
h2	&	&9\,803	&	&1.04	&1.26	&1.31	&	&1.6	&9.8	&20.2	&19	&0	&	&1\,460	&1.033	&	&96.0	&1.9	&2.1	&	&0.4	&	&4.6	\\
lusearch	&	&3\,461	&	&1.02	&1.04	&1.01	&	&154.8	&1.0	&1.4	&21	&5	&	&5\,058	&1.022	&	&99.5	&0.2	&0.3	&	&0.0	&	&0.4	\\
tomcat	&	&5\,203	&	&1.01	&1.10	&1.09	&	&23.2	&2.3	&3.9	&14	&0	&	&2\,426	&1.013	&	&99.4	&0.5	&0.2	&	&0.0	&	&0.5	\\
\mymidrule
avrora	&	&3\,822	&	&1.02	&1.02	&1.06	&	&15.8	&1.1	&1.5	&23	&1	&	&\,479	&0.997	&	&94.9	&0.1	&5.0	&	&0.2	&	&4.6	\\
batik	&	&1\,880	&	&0.99	&1.00	&1.01	&	&1.1	&20.7	&24.5	&50	&0	&	&3\,511	&1.022	&	&42.9	&2.6	&54.5	&	&6.6	&	&52.6	\\
biojava	&	&15\,695	&	&1.00	&1.00	&1.00	&	&4.5	&1.3	&4.5	&10	&0	&	&1\,596	&1.021	&	&100.0	&0.0	&0.0	&	&0.0	&	&0.0	\\
eclipse	&	&14\,878	&	&1.00	&1.04	&1.05	&	&1.5	&5.4	&12.4	&29	&0	&	&\,845	&1.039	&	&93.8	&2.0	&4.3	&	&1.0	&	&7.8	\\
fop	&	&\,933	&	&1.00	&1.02	&1.02	&	&5.2	&2.9	&4.4	&25	&0	&	&\,776	&1.038	&	&97.7	&0.0	&2.3	&	&0.1	&	&2.7	\\
graphchi	&	&10\,935	&	&1.00	&1.01	&1.00	&	&4.1	&1.7	&4.3	&24	&0	&	&\,512	&0.977	&	&95.4	&0.0	&4.6	&	&0.0	&	&1.7	\\
h2o	&	&4\,365	&	&1.00	&1.00	&1.00	&	&0.5	&12.6	&13.3	&0	&0	&	&\,70	&1.046	&	&100.0	&0.0	&0.0	&	&0.0	&	&0.1	\\
jython	&	&5\,340	&	&1.00	&1.01	&1.00	&	&1.9	&2.7	&3.1	&0	&0	&	&\,219	&1.041	&	&100.0	&0.0	&0.0	&	&0.0	&	&0.0	\\
luindex	&	&6\,602	&	&1.01	&1.01	&1.02	&	&6.6	&1.1	&1.6	&19	&0	&	&\,204	&1.015	&	&99.6	&0.3	&0.1	&	&0.0	&	&0.4	\\
pmd	&	&2\,071	&	&1.03	&1.02	&1.02	&	&5.1	&13.5	&20.1	&33	&0	&	&9\,296	&1.028	&	&93.1	&0.0	&6.9	&	&0.6	&	&10.6	\\
sunflow	&	&2\,894	&	&1.02	&1.02	&1.01	&	&58.7	&1.5	&2.0	&20	&0	&	&6\,202	&1.010	&	&97.9	&0.3	&1.8	&	&0.0	&	&2.5	\\
xalan	&	&\,920	&	&0.82	&0.94	&0.91	&	&210.6	&1.3	&1.7	&20	&22	&	&9\,197	&1.017	&	&94.8	&0.8	&4.4	&	&0.2	&	&9.8	\\
zxing	&	&\,829	&	&0.99	&1.01	&1.00	&	&10.3	&1.4	&2.0	&27	&0	&	&\,429	&0.968	&	&99.4	&0.3	&0.2	&	&0.0	&	&0.4	\\\mymidrule    
																								
\multicolumn{1}{r}{\textbf{min}}	&	&	&	&\emph{0.82}	&\emph{0.94}	&\emph{0.91}	&	&\emph{0.5}	&\emph{1.0}	&\emph{1.4}	&\emph{0}	&\emph{22}	&	&\emph{70}	&\emph{0.968}	&	&\emph{42.9}	&\emph{0.0}	&\emph{0.0}	&	&\emph{0.0}	&	&\emph{0.0}	\\
\multicolumn{1}{r}{\textbf{max}}	&	&	&	&\emph{1.04}	&\emph{1.26}	&\emph{1.31}	&	&\emph{210.6}	&\emph{20.7}	&\emph{24.5}	&\emph{50}	&\emph{2}	&	&\emph{9\,296}	&\emph{1.046}	&	&\emph{100.0}	&\emph{2.6}	&\emph{54.5}	&	&\emph{6.6}	&	&\emph{52.6}	\\
\multicolumn{1}{r}{\textbf{mean}}	&	&	&	&\emph{1.00}	&\emph{1.03}	&\emph{1.03}	&	&\emph{29.9}	&\emph{5.0}	&\emph{7.5}	&\emph{21}	&\emph{2}	&	&\emph{2\,511}	&\emph{1.017}	&	&\emph{94.3}	&\emph{0.6}	&\emph{5.1}	&	&\emph{0.5}	&	&\emph{5.9}	\\
\multicolumn{1}{r}{\textbf{geomean}}	&	&	&	&\emph{1.00}	&\emph{1.03}	&\emph{1.03}	&	&\emph{6.7}	&\emph{3.0}	&\emph{4.7}	&	&	&	&\emph{1\,094}	&\emph{1.016}	&	&\emph{92.9}	&\emph{0.1}	&\emph{0.0}	&	&\emph{0.0}	&	&\emph{1.1}	\\\bottomrule

\end{tabular}
}

%% file: discussion.tex
\label{sec:discussion}

\section{Discussion and Threats to Validity}
\label{sec:threats}

\subsubsection*{OpenJDK 11}
\label{sec:threats:openjdk11}

Building an entirely new garbage collector in OpenJDK is a major undertaking.  We implement \lxr in OpenJDK 11 because it is a long term service (LTS) release, including significant back-ported improvements to Shenandoah and ZGC.  On 2021-09-14, JDK 17 reached \emph{General Availability} status and is now the long-term support (LTS) release, following JDK 11~\cite{Oracle:21}.  We plan to port our work to JDK 17.   Because of the substantial differences between JDK 11 and JDK 17~\cite{Oracle:21b}, it is not possible to compare our collector and the JDK 17 collectors directly.  The JDK 17 version of ZGC removes its limitation of operating only in large heaps.  To validate our most striking findings on Lucene, we ran Shenandoah from OpenJDK 17 and found that its performance was essential identical to the results with OpenJDK 11---even with the many advances included in JDK 17, it still exhibits the same pathologies we report here.

C4 is implemented in Platform Prime (formerly Zing) JDK, not OpenJDK~\cite{TIW:11}.    Apples-to-apples collector comparisons require  using  the same runtime.   However, the design decisions that we critique and respond to in this paper are common to C4, Shenandoah, and ZGC.   The notable advantage of C4 is its generational collection.  For highly allocating workloads such as \textsf{lusearch}, the non-generational Shenandoah and ZGC are at a  disadvantage with respect to C4, G1, and \lxr, which all optimize for generational behavior.  

\subsubsection*{Compressed Pointers, Weak References, and Class Unloading}
\label{sec:threats:cp}

{We have not yet implemented compressed pointers so we disable it in our evaluations~\cite{LA:05}.   Although profitable, we believe it is orthogonal to \lxr and do not expect its absence to change our findings.  We have not completed the implementation of weak references or class unloading, so we also disable them on all systems. Turning off compressed pointers increased minimum heap sizes for most benchmarks and together with the other disabled  features slowed G1 down by just 2.6\% at a $3\times$ heap size.}

\subsubsection*{Implementation and Workloads}
\label{sec:threats:prod}

{Our collector is new so has neither had the benefit of years of heuristic tuning, testing, or optimization by a production team. We believe that our ability to execute on 17 substantial workloads from the recent Chopin development branch of DaCapo demonstrates  robustness and completeness.} 

\section{Conclusion}
\label{sec:future}



Since 2004, G1's region-based design with concurrent tracing and strict evacuation has dominated production garbage collectors~\cite{TIW:11,Liden:17,FKD+:16,ZB:20}.  This paper identifies fundamental limitations in timeliness due to full heap tracing in all systems and in the concurrent copying approach in C4, Shenandoah, and ZGC.  Although they deliver very short pause times, short pauses do not always translate into low-latency for request-driven time critical workloads. We introduce \lxr, which takes an entirely different approach. It uses brief stop-the-world pauses and reference counting to promptly reclaim most memory without copying. It defragments the heap with limited opportunistic copying. It concurrently reclaims mature objects, identifying dead objects in cycles and dead objects with stuck reference counts. {The result is a collector with moderate pause times that delivers low application latency and high throughput.  Our results are robust to heap size and microarchitecture. These  initial results demonstrate that this new design delivers performance without requiring additional hardware or  memory.}   We hope that this work will provoke a rethink of modern collector design and a reinvigoration of GC research.